\documentclass[12pt,preprint]{aastex}

\def\dex#1{\hbox{$10^{#1}$}}
\def\tdex#1{\hbox{$\times10^{#1}$}}
\def\deg{$^\circ$}
\def\cmm#1{\hbox{${\rm cm^{-#1}}$}}
\def\kms{km\,s$^{-1}$}

\def\Lya{Ly$\alpha$}
\def\HI{\protect\ion{H}{1}}
\def\CII{\protect\ion{C}{2}}
\def\CIII{\protect\ion{C}{3}}
\def\NII{\protect\ion{N}{2}}
\def\NIII{\protect\ion{N}{3}}
\def\NV{\protect\ion{N}{5}}
\def\OI{\protect\ion{O}{1}}
\def\OIV{\protect\ion{O}{4}}
\def\OVI{\protect\ion{O}{6}}
\def\NeVIII{\protect\ion{Ne}{8}}

\def\SiII{\protect\ion{Si}{2}}

\def\SIV{\protect\ion{S}{4}}

\def\FeIII{\protect\ion{Fe}{3}}

\def\HST{{\it HST}}
\def\FOS{{\it FOS}}

\def\STIS{{\it STIS}}

\def\MAST{{\it MAST}}
\def\NHI{$N$(\HI)}
\def\NHITWcm{$N$(\HI; 21cm)}
\def\NHIHVC{$N$(\HI; HVC)}
\def\NHILAB{$N$(\HI; LAB)}
\def\NHIGB{$N$(\HI; 140')}
\def\NHIGBT{$N$(\HI; GBT)}

\def\NHILya{$N$(\HI; \Lya)}
\def\Nav{$\bar N$}
\def\NoNav{$N$/$\bar N$}
\def\Davg{$<$$D$$>$}
\def\Ravg{$<$$R$$>$}
\def\ref{\par}

\def\Sobs{2}

\def\Sfitmethod{2.2}
\def\Stocm{2.3}
\def\SGBT{2.4}
\def\Scorrect{3}
\def\Sresults{4}

\def\Sdataqual{4.3}

\def\Sdisc{5}
\def\Sttest{5.1}

\def\Fmap{1}
\def\Fcompare{2}
\def\Fnosignal{3}
\def\Fexample{4}
\def\Fspectra{5}
\def\Fscatter{6}
\def\Fratio{7}
\def\Ftheor{8}
\def\Tobs{1}
\def\Tres{2}
\def\TFOS{3}
\def\Tstat{4}

\begin{document}

\title{Measuring turbulence in the ISM by comparing $N$(\HI;\Lya) and $N$(\HI;21-cm)}

\author{%
Bart P.\ Wakker\altaffilmark{1},
Felix J.\ Lockman\altaffilmark{2},
Jonathan M.\ Brown\altaffilmark{1,3}
\altaffiltext{1}{Department of Astronomy, University of Wisconsin, Madison, WI 53706; jbrown3@wisc.edu, wakker@astro.wisc.edu}
\altaffiltext{2}{National Radio Astronomy Observatory, Green Bank, WV 24944, jlockman@nrao.edu}
\altaffiltext{3}{Presently at: Hubert H.\ Humphrey Institute of Public Affairs, Minneapolis, MN 55455, brow3019@umn.edu}
}

\begin{abstract}

We present a study of the small-scale structure of the interstellar medium in
the Milky Way. We used HST STIS data to measure $N$(\HI) in a pencil-beam toward
59 AGNs and compared the results with the values seen at 9\arcmin--36\arcmin\
resolution in the same directions using radio telescopes (GBT, Green Bank 140-ft
and LAB survey). The distribution of ratios $N$(\Lya)/$N$(\HI) has an average of
1 and a dispersion of about 10\%. Our analysis also revealed that spectra from
the Leiden-Argentina-Bonn (LAB) all-sky \HI\ survey need to be corrected, taking
out a broad gaussian component (peak brightness temperature 0.048\,K, FWHM
167~\kms, and central velocity $-$22~\kms). The column density ratios have a
distribution showing similarities to simple descriptions of hierarchical
structure in the neutral ISM, as well as to a more sophisticated 3D MHD
simulation. From the comparison with such models, we find that the sonic Mach
number of the local ISM should lie between 0.6 and 0.9. However, none of the
models yet matches the observed distribution in all details, but with many more
sightlines (as will be provided by COS) our approach can be used to constrain
the properties of interstellar turbulence.

\end{abstract}

\keywords{
ISM: clouds,
ISM: general,
ISM: structure,
turbulence
}

\section{Introduction}
\par A fundamental aspect of understanding the structure of the interstellar
medium (ISM) involves the origin of the small-scale structure that is observed.
A number of formulations provide a framework for analyzing such structure, and
explaining its creation and distribution. These formulations include those of
Houlahan \& Scalo (1992), who developed a hierarchical tree structure,
describing clouds as a series of partitions. Falgarone et al.\ (1991) explored
fractal structure and showed that this description is applicable to molecular
clouds. Vogelaar \& Wakker (1994) used fractal structure in an attempt to
describe the structure of high-velocity clouds. Fractal structure may arise
naturally from the density statistics of interstellar turbulence. More recently,
Lazarian \& Pogosyan (2000, 2004, 2006, 2008) developed techniques for comparing
observations of velocity and density structure in the ISM with statistics
derived from theories of turbulence. Kowal et al.\ (2007) used these ideas to
construct 3-D MHD models of the ISM that predict the spectrum of density
fluctuations, while Burkhart et al.\ (2009) studied how statistical measures can
be used to connect these models to observations. In this paper, we look at the
column density distribution of neutral hydrogen, trying to compare measurements
made at different angular resolutions, which allows us to apply one of the
measures discussed by Burkhart et al.\ (2009).
\par Galactic \HI\ column densities can be measured using 21-cm radio
observations or by using \Lya\ absorption-line spectra. 21-cm observations are
made with single-dish or interferometer radiotelescopes. Single-dish telescopes
have a large beam size, typically 36\arcmin\ for all-sky surveys, and
9\arcmin--21\arcmin\ for more targeted observations. Interferometers produce
beams of less than 2\arcmin, although for Galactic gas it is necessary to
combine this with single-dish observations to derive an accurate total column
density. With \Lya\ observations of ultra-violet bright background targets,
however, one measures \NHI\ in a very small (sub arcsecond) area centered on the
background target. If the background target is a galactic disk star, the \Lya\
absorption is caused only by the \HI\ in front of the star, whereas the 21-cm
observations measure \HI\ both in front of and behind the star. When observing
AGNs, both \Lya\ and 21-cm observations mesaure all \HI\ in the line of sight.
Some halo stars may also be sufficiently high above the galactic plane to lie
above most of the \HI. Thus, valid comparisons between \NHILya\ and \NHITWcm\
can only be made in the directions of AGNs or distant halo stars.
\par Precise measurements of \NHI\ are also necessary to derive abundances of
heavy elements in the interstellar medium (ISM). Usually, 21-cm data are used to
derive the \HI\ reference column density, which is necessary whenever absorption
lines show different components. However, the metal-line absorption is produced
only by the ions in the small beam toward the AGN, while the 21-cm data average
\NHI\ across the radiotelescope beam. Thus, there is a question as to whether it
is correct to use the 21-cm data to derive \NHI. Do \Lya\ measurements and 21-cm
measurements in fact give consistent results? If they do not, what is the reason
for this difference?
\par Hobbs et al.\ (1982) were the first to directly compare \NHI\ measured
using \Lya\ absorption lines (from Copernicus data) to \NHITWcm, measured using
the 140-ft Green Bank telescope (21\arcmin\ beam). Although they did not present
errors on their meaurements, the ratios they found varied from about 0.9 to 3.9,
with the outliers for the stars closest to the plane. For the three stars above
most of the Galactic \HI\ layer ($z$$>$2~kpc) the ratios were given as 0.9, 0.9
and 1.1.
\par This was followed by a study by Lockman et al.\ (1986a), who used {\it
International Ultraviolet Explorer} (IUE) spectra to measure \NHILya\ toward 45
stars, by fitting the profiles of the damping wings. The resulting errors in
$N$(\HI) were estimated to be about 0.1 dex (25\%). These values were again
compared to 140-ft Green Bank Telescope 21-cm data, which were corrected for
stray radiation. They also used a spin temperature of 75\,K to correct the
column densities for optical depth effects, though in most cases this correcton
is small (a few percent). For the stars closest to the plane ($z$$<$1~kpc), the
ratio \NHILya/\NHITWcm\ tends to be $<$1, because not all of the \HI\ in the
sightline is seen in absorption. For the six stars at $z$$>$1.5 kpc the average
ratio was about 1, to within the errors, although the typical error in each
ratio was about 0.3.
\par Savage et al.\ (2000) revisited this comparison in the directions of 14
QSOs, using data from the G130H grating in the Faint Object Spectrograph (\FOS)
on the Hubble Space Telescope (\HST). Unlike the Copernicus and IUE data, the
\FOS\ spectra had low resolution (230~\kms). Savage et al.\ (2000) compared
these masurements to 21-cm data from the Green Bank 140-ft telescope. For ten of
the QSO spectra it was possible to correct for geocoronal emission, and for
these Savage et al.\ (2000) found that the ratio \NHILya/\NHITWcm\ had values in
the range 0.62 to 0.91 with errors of about 0.1.
\par Savage et al.\ (2000) suggested two possible origins for differences
between the column densities measured from \Lya\ absorption and 21-cm emission.
First, differences might arise from a combination of systematic and random
errors in the \Lya\ and 21-cm observations. In the \Lya\ observations,
systematic errors can be produced by uncertain continuum placement, geocoronal
\HI\ emission removal, the spectrograph background and scattered light
correction, interfering QSO and IGM absorption, and detector fixed pattern
noise. For the \FOS\ dataset of Savage et al.\ (2000) all of these effects were
present. Using data at higher spectral resolution removes all of the trouble
associated with geocoronal \HI, IGM absorption and fixed pattern noise, while it
greatly reduces the other problems. In the 21-cm data, systematic errors can be
created by the absolute calibration of the radio telescope, baseline fitting and
the stray-radiation correction. Alternatively, the differences in \Lya\ and
21-cm column densities could be caused by the structure of the ISM. If there are
small bright spots embedded in a smoother background, the 21-cm data will
include these, but a random sightline to an AGN has a high probability of
missing the brighter spots. To study such effects requires a large sample of
sightlines.

\par In this paper we analyze 59 sightlines using higher-resolution and higher
S/N \Lya\ spectra than used by Hobbs et al.\ (1982), Lockman et al.\ (1986b) and
Savage et al.\ (2000). We compared all these measurements to the column
densities found in the LAB survey (Kalberla et al.\ 2005), as well as to the
Lockman \& Savage (1995) Green Bank 140-ft data that is available in many
directions. In addition, we obtained new 21-cm data with the Green Bank
Telescope (GBT) toward 35 AGNs. In Sect.~\Sobs\ we describe the \Lya\ and 21-cm
datasets that we used, as well as the method used to derive \NHILya. During our
analysis, we discovered the presence of a spurious, broad underlying component
in the LAB and 140-ft spectra. We show this in Sect.~\Scorrect. After removing
this component, we find the results that are presented in Sect.~\Sresults\ and
discussed in Sect.~\Sdisc.

\begin{figure}\plotfiddle{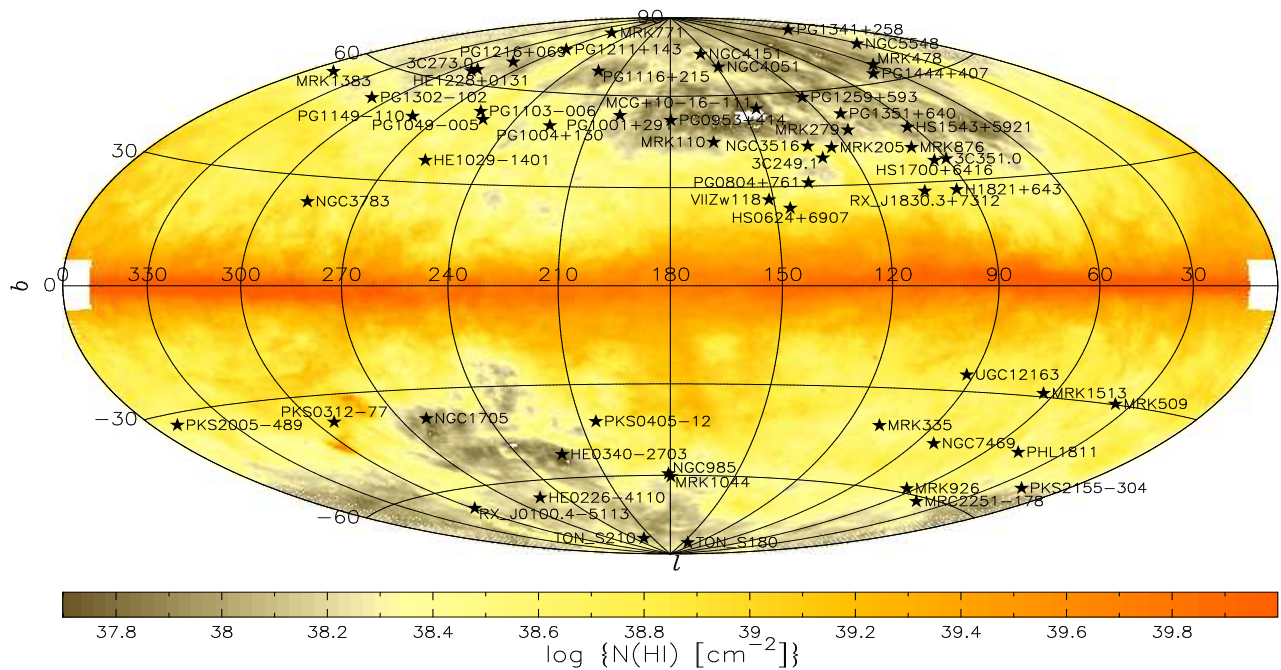}{0in}{270}{250}{480}{0}{-350}\figurenum{\Fmap}
\caption{All-sky map of \NHI\ integrated between $-$100 and 100~\kms, using the
Leiden-Argentina Bonn survey (Kalberla et al.\ 2005), with the column density
scale on the bottom. The directions to our 59 targets are shown by the stars and
labels.}
\end{figure}

\begin{deluxetable}{lrrrllrrrc}
\tabletypesize{\footnotesize}
\tablenum{1} \tablewidth{0pt}
\tablecolumns{10}
\tablecaption{Observational data}
\tablehead{%
\colhead{Object} &\colhead{Lon}    &\colhead{Lat}    &\colhead{Type} &\colhead{Obs. ID.} &\colhead{Grating} &\colhead{$\lambda_{\rm min}$} &\colhead{$\lambda_{\rm max}$} &\colhead{T$_{\rm exp}$} &\colhead{GBT} \\
         &\colhead{[$^\circ$]} &\colhead{[$^\circ$]} &\colhead{}     &\colhead{}         &\colhead{}        &\colhead{[\AA]}               &\colhead{[\AA]}               &\colhead{[ks]}          &              \\
\colhead{(1)}    &\colhead{(2)}    &\colhead{(3)}    &\colhead{(4)}  &\colhead{(5)}      &\colhead{(6)}     &\colhead{(7)}                 &\colhead{(8)}                 &\colhead{(9)}           &\colhead{(10)}\\
}
\startdata
3C249.1          &      130.39     &       38.55     &     QSO       & O6E1/24$-$30        &   E140M          & 1140  &  1729   & 68.8 & Y \\ 
3C273.0          &      289.95     &       64.36     &     QSO       & O5D3/01$-$02        &   E140M          & 1140  &  1709   & 18.7 & Y \\ 
3C351.0          &       90.08     &       36.38     &     QSO       & O579/01$-$04        &   E140M          & 1140  &  1709   & 77.0 & Y \\ 
H1821+643        &       94.00     &       27.42     &     QSO       & O5E7/03$-$05        &   E140M          & 1140  &  1709   & 50.9 & Y \\ 
HE0226$-$4110      &      253.94     &      -65.77     &     QSO       & O6E1/07-11        &   E140M          & 1143  &  1729   & 43.8 &   \\ 
HE0340$-$2703      &      222.68     &      -52.12     &     QSO       & O8EI/03           &   G140M          & 1194  &  1250   &  4.9 &   \\ 
HE1029$-$1401      &      259.33     &       36.52     &     QSO       & O4EC/05           &   G140M          & 1194  &  1248   &  4.1 &   \\ 
HE1228+0131      &      291.26     &       63.66     &     QSO       & O56A/01$-$02        &   E140M          & 1140  &  1729   & 27.2 &   \\ 
HS0624+6907      &      145.71     &       23.35     &     QSO       & O6E1/12$-$16        &   E140M          & 1140  &  1729   & 62.0 & Y \\ 
HS1543+5921      &       92.40     &       46.36     &     QSO       & O8MR/01           &   G140M          & 1194  &  1250   & 25.3 &   \\ 
HS1700+6416      &       94.40     &       36.16     &     QSO       & O4SI/01$-$05        &   E140M          & 1140  &  1729   & 99.4 &   \\ 
MCG+10$-$16-111    &      144.21     &       55.08     &     Sey       & O5EW/02           &   G140M          & 1194  &  1249   & 19.5 & Y \\ 
MRC2251$-$178      &       46.20     &      -61.33     &     QSO       & O4EC/03           &   G140M          & 1194  &  1249   &  6.0 & Y \\ 
Mrk110           &      165.01     &       44.36     &     Sey       & O4N3/52           &   G140M          & 1194  &  1249   &  2.2 & Y \\ 
Mrk205           &      125.45     &       41.67     &     Sey       & O62Q/03$-$05        &   E140M          & 1140  &  1729   & 62.1 & Y \\ 
Mrk279           &      115.04     &       46.86     &     Sey       & O6JM/01           &   E140M          & 1140  &  1709   & 13.2 &   \\ 
                 &                 &                 &               & O8K1/01$-$05        &   E140M          & 1140  &  1709   & 52.8 &   \\ 
Mrk335           &      108.76     &      $-$41.42     &     Sey       & O8N5/04-05        &   E140M          & 1140  &  1729   & 15.2 & Y \\ 
Mrk478           &       59.24     &       65.03     &     Sey       & O4EC/14           &   G140M          & 1194  &  1249   &  7.6 & Y \\ 
Mrk509           &       35.97     &      $-$29.86     &     Sey       & O6AP/01           &   E140M          & 1140  &  1709   &  7.6 & Y \\ 
Mrk771           &      269.44     &       81.74     &     Sey       & O4N3/05           &   G140M          & 1194  &  1249   &  2.0 & Y \\ 
                 &                 &                 &               & O4EC/07           &   G140M          & 1194  &  1249   &  5.8 &   \\ 
Mrk876           &       98.27     &       40.38     &     Sey       & O8NN/01$-$02        &   E140M          & 1140  &  1729   & 29.2 & Y \\ 
Mrk926           &       64.09     &      $-$58.76     &     Sey       & O4EC/12           &   G140M          & 1194  &  1249   &  3.9 &   \\ 
Mrk1044          &      179.69     &      $-$60.48     &     Sey       & O8K4/01           &   G140M          & 1194  &  1250   &  2.4 & Y \\ 
Mrk1383          &      349.22     &       55.13     &     Sey       & O8PG/01$-$02        &   E140M          & 1140  &  1729   & 19.2 & Y \\ 
Mrk1513          &       63.67     &      $-$29.07     &     Sey       & O4EC/10           &   G140M          & 1194  &  1249   &  7.3 & Y \\ 
NGC985           &      180.84     &      $-$59.49     &     Sey       & O4EC/11           &   G140M          & 1194  &  1249   &  3.7 & Y \\ 
NGC1705          &      261.08     &      $-$38.74     &     Sey       & O58N/01-02        &   E140M          & 1140  &  1709   & 17.1 &   \\ 
NGC3516          &      133.24     &       42.40     &     Sey       & O57B/02           &   E140M          & 1140  &  1729   &  5.5 &   \\ 
NGC3783          &      287.46     &       22.95     &     Sey       & O57B/01           &   E140M          & 1140  &  1729   &  5.4 &   \\ 
                 &                 &                 &               & O63M/01$-$17        &   E140M          & 1140  &  1729   & 87.0 &   \\ 
                 &                 &                 &               & O63M/53           &   E140M          & 1140  &  1729   &  4.9 &   \\ 
NGC4051          &      148.88     &       70.09     &     Sey       & O5F0/01           &   E140M          & 1140  &  1729   & 10.3 &   \\ 
NGC4151          &      155.08     &       75.06     &     Sey       & O578/01           &   E140M          & 1140  &  1729   &  5.5 &   \\ 
                 &                 &                 &               & O5KT/02,53,54     &   E140M          & 1140  &  1729   &  6.6 &   \\ 
                 &                 &                 &               & O61L/01           &   E140M          & 1140  &  1729   &  9.4 &   \\ 
                 &                 &                 &               & O6JB/01           &   E140M          & 1140  &  1729   &  7.6 &   \\ 
NGC5548          &       31.96     &       70.50     &     Sey       & O4LL/01           &   E140M          & 1140  &  1729   &  4.8 & Y \\ 
                 &                 &                 &               & O6JD/01$-$02        &   E140M          & 1140  &  1729   & 15.3 &   \\ 
                 &                 &                 &               & O6KW/01$-$04        &   E140M          & 1140  &  1729   & 52.2 &   \\ 
NGC7469          &       83.10     &      $-$45.47     &     Sey       & O6BN/01           &   E140M          & 1140  &  1729   & 13.0 & Y \\ 
                 &                 &                 &               & O8N5/01$-$02        &   E140M          & 1140  &  1729   & 22.8 &   \\ 
PG0804+761       &      138.28     &       31.03     &     QSO       & O4N3/01           &   G140M          & 1194  &  1249   &  2.4 & Y \\ 
                 &                 &                 &               & O4EC/06           &   G140M          & 1194  &  1249   &  4.9 & Y \\ 
PG0953+414       &      179.79     &       51.71     &     QSO       & O4X0/01$-$02        &   E140M          & 1140  &  1729   & 25.5 & Y \\ 
PG1001+291       &      200.08     &       53.21     &     QSO       & O6E1/17$-$23        &   E140M          & 1140  &  1729   & 48.4 & Y \\ 
PG1004+130       &      225.12     &       49.12     &     QSO       & O5EW/03           &   G140M          & 1194  &  1249   &  5.2 &   \\ 
PG1049$-$005       &      252.28     &       49.88     &     QSO       & O4N3/03           &   G140M          & 1194  &  1249   &  1.5 &   \\ 
PG1103$-$006       &      256.66     &       52.30     &     QSO       & O4N3/04           &   G140M          & 1194  &  1249   &  1.4 &   \\ 
PG1116+215       &      223.36     &       68.21     &     QSO       & O5A3/01$-$02        &   E140M          & 1139  &  1709   &  6.6 &   \\ 
                 &                 &                 &               & O5E7/01$-$02        &   E140M          & 1140  &  1709   & 19.9 &   \\ 
PG1149$-$110       &      280.47     &       48.89     &     Sey       & O5EW/05           &   G140M          & 1194  &  1250   &  8.1 &   \\ 
PG1211+143       &      267.55     &       74.31     &     Sey       & O61Y/01$-$08        &   E140M          & 1140  &  1729   & 42.5 & Y \\ 
PG1216+069       &      281.07     &       68.14     &     QSO       & O6E1/31$-$39        &   E140M          & 1140  &  1729   & 69.8 & Y \\ 
PG1259+593       &      120.56     &       58.05     &     QSO       & O63G/05$-$11        &   E140M          & 1143  &  1729   & 95.8 &   \\ 
PG1302$-$102       &      308.59     &       52.16     &     QSO       & O5BU/01,02,61     &   E140M          & 1140  &  1729   & 22.1 & Y \\ 
PG1341+258       &       28.71     &       78.15     &     QSO       & O5EW/01           &   G140M          & 1194  &  1250   &  6.9 & Y \\ 
PG1351+640       &      111.89     &       52.02     &     Sey       & O4EC/54           &   G140M          & 1194  &  1248   & 14.7 & Y \\ 
PG1444+407       &       69.90     &       62.72     &     QSO       & O6E1/01$-$06        &   E140M          & 1140  &  1729   & 48.6 & Y \\ 
PHL1811          &       47.47     &      $-$44.82     &     QSO       & O8D9/01-04        &   E140M          & 1140  &  1729   & 33.9 & Y \\ 
PKS0312$-$77       &      293.44     &      -37.55     &     QSO       & O65T/01,02,13     &   E140M          & 1140  &  1729   &  8.4 &   \\ 
PKS0405$-$12       &      204.93     &      -41.76     &     QSO       & O55S/01,02        &   E140M          & 1139  &  1729   & 27.2 & Y \\ 
PKS2005$-$489      &      350.37     &      -32.60     &     BLLac     & O4EC/09           &   G140M          & 1194  &  1249   &  6.1 &   \\ 
PKS2155$-$304      &       17.73     &      -52.25     &     BLLac     & O5BY/01-02        &   E140M          & 1139  &  1729   & 28.5 &   \\ 
RX J0100.4$-$5113  &      299.48     &      -65.84     &     Sey       & O8P8/02           &   G140M          & 1194  &  1250   &  2.3 &   \\ 
RX J1830.3+7312  &      104.04     &       27.40     &     Sey       & O5EW/09           &   G140M          & 1194  &  1249   &  5.8 & Y \\ 
Ton S180         &      139.00     &      $-$85.07     &     Sey       & O4EC/02           &   G140M          & 1194  &  1249   &  4.1 & Y \\ 
Ton S210         &      224.97     &      $-$83.16     &     QSO       & O6L0/01-02        &   E140M          & 1140  &  1709   & 12.3 &   \\ 
UGC12163         &       92.14     &      $-$25.34     &     Sey       & O5IT/05           &   E140M          & 1140  &  1709   & 10.3 &   \\ 
VIIZw118         &      151.36     &       25.99     &     Sey       & O4EC/13           &   G140M          & 1194  &  1249   &  9.5 & Y \\ 
\enddata
\tablecomments{
Col.\ (1): Object name. Cols.\ (2,3): Galactic longitude and latitude. Col.\ (4):
Object type; QSO is a quasar and Sey is a Seyfert galaxy. Col.\ (5):
Observational program identification datasets with exposure identifications.
Col.\ (6): Grating used for observation; either \HST\ \STIS$-$G140M or \HST\
\STIS$-$E140M. Cols.\ (7,8): Minimum and maximum wavelength of
observational wavelength spanned. Col.\ (9): Total exposure time. Col.\ (10):
``Y'' means there is a spectrum taken with the Green Bank Telescope.
}
\end{deluxetable}

\section{Observations}

\subsection{\HST\ data}
\par We used Hubble Space Telescope (\HST) observations of AGNs that were
observed at sufficient spectral resolution to (mostly) resolve the interstellar
and intergalactic absorption lines. This includes observations using the G140M
grating and E140M echelle in the {\it Space Telescope Imaging Spectrograph}
(\STIS). G140M spectra cover a wavelength interval about 55~\AA\ wide at
30~\kms\ resolution, while E140M spectra range from 1140 to 1710~\AA\ with a
resolution of 6.5~\kms. The calibrated \HST\ data were retrieved from the {\it
Multimission Archive at STScI} (\MAST) server. We do not look at the many
targets observed with the \STIS-G140L grating, since the interstellar lines near
1200 and 1206~\AA\ in the damping profile are not fully resolved in such
low-resolution (300~\kms) data, nor are low-redshift intergalactic \Lya\ or
high-redshift intergalactic metal lines. When unresolved, such lines change the
apparent continuum, making it more difficult to obtain reliable results. We list
the sample of 59 AGNs in Table~\Tobs. The locations of these targets on the sky
are shown in Fig.~\Fmap.

\subsection{Fitting Method}
\par To measure \NHI, the continuum-reconstruction method was used, which works
as follows. First, one assumes an \HI\ column density and a linewidth, which
yields an optical depth ($\tau$) profile. Multiplying the observed spectrum with
$\exp(+\tau)$ then gives a ``reconstructed'' continuum, which is assumed to be
smooth as well as continuous with the parts of the spectrum where the continuum
is observed directly. The \HI\ column density is then varied until this is
actually the case. We now describe this in more detail.

\par Before applying this method, we first increased the signal to noise (S/N)
ratio of the spectra by binning the E140M data by 15 pixels (to 48~\kms\ or 7
resolution elements), and the G140M data by 3 pixels (to 40~\kms\ or 1.3
resolution elements). Then we defined a continuum ($C$) by selecting regions
free of absorption and emission in a window about 50~\AA\ wide around the \Lya\
line and fitting a Legendre polynomial of order up to 4 through these selected
points, using the method of Sembach \& Savage (1992). To ensure the continuity
and smoothness of the reconstructed continuum, some of these regions are placed
so that the final fit is made through the reconstructed continuum in regions
where there is \Lya\ absorption.
\par Next, we created a reconstructed continuum starting with the \NHI\ value
obtained from the 21-cm brightness temperature. The velocity of the \HI\ profile
was also determined from the 21-cm \HI\ emission data toward the sightline (see
Sect.~\Stocm\ below). Using the velocity and column density, we multiplied the
observed flux ($F$) by $\exp(+\tau)$, where $\tau$ is the optical depth of the
\Lya\ line, including its damping wings. We note that the width of the central
gaussian part of the absorption profile is not important for the shape of the
damping wings.
\par Finally, we made a fit to the continuum by changing the column density
value until the fitted continuum a) looked smooth and b) minimized the
difference between the fit and the reconstruction. The minimization is done by
calculating a reduced $\chi^2$ as:
$$\chi^2\ =\ {1\over n}\ \Sigma \left({ C - F\ e^\tau \over \delta F\ e^\tau}\right)^2,
$$ where $C$ is the fit to the reconstucted continuum, $F$ is the observed flux,
$\delta F$ is the error in the flux, and $n$ is the number of independent
pixels. This calculation is only done in selected regions of the spectrum, which
are different from those used to define the continuum. The selected regions are
those that a) are within about 10~\AA\ of the central wavelength of \Lya\
(1215.67~\AA), b) are free of intergalactic or interstellar absorption lines, c)
have the optical depth of the \Lya\ line $<$1.5, and d) have an S/N ratio
($F$/$\delta F$) $>$2. The range of \NHI\ values at which $\chi^2$=$\chi^2_{\rm
min}$+1 determines the error in each \NHI\ result.
\par Some sightlines contain high-velocity \HI\ components. In such cases, we
applied a procedure to determine a systematic error in \NHI\ for the
low-velocity components. To do this, we systematically varied the column density
of one component, while fitting the other. Specifically, for sightlines with a 
HVC having \NHI\ between $\sim$2\tdex{19} and \dex{20}\,\cmm2, we fixed \NHIHVC\
at the nominal 21-cm value, as well as at nominal$\times$1.5 and nominal/1.5,
and then fitted the column density of the low-velocity component for each of
these three HVC component values. The resulting range of \NHI\ values for the
low-velocity component then gave an estimate of the uncertainty associated with
having the high-velocity component present. For sightlines containing high- and
low-velocity components with similar strengths, the components were varied by
20\% instead of 50\%, since the result is better constrained in this case. We
then executed the procedure of fixing one component and fitting the other for
both components separately, while defining a region to calculate $\chi^2$ on
only one side of the \Lya\ absorption line. In most of these case we decided in
the end that the systematic error in the final values was too large to use the
results in the analysis of column density ratios.

\subsection{21-cm data}
\par We used three sources of 21-cm data. For 35 of the targets that have \Lya\
spectra we obtained new data with the Green Bank Telescope (GBT) at an angular
resolution of 9\farcm1 (see below for more details on these observations). In
167 directions (42 overlapping with the \Lya\ sample), we also used the spectra
of Lockman \& Savage (1995), which were taken with the Green Bank 140-ft
(21\arcmin\ beam). For each of the directions in the merged \Lya+GBT+140\arcmin\
list, we interpolated between the grid points of the Leiden-Argentina-Bonn
survey (Kalberla et al.\ 2005), a whole-sky survey done at 1.3~\kms\ velocity
resolution with a 36\arcmin\ beam on a 0\fdg5$\times$0\fdg5 grid.
For each object, the 21-cm column density, \NHITWcm, was obtained by integrating
the brightness temperature over a specified velocity interval and converting
this to a column density using $$
N(H\,I)\ =\int\ 1.823\tdex{18}\ T_s\ \ln\left( {T_s \over T_s-T_{\rm B} } \right)\ dv.
$$

\par We analyze the highest $T_B$ half of each spectrum with $T_s$=135K and the
faintest $T_B$ half with $T_s$=5000\,K and then add the two values.

This is justified by the
fact that Heiles \& Troland (2003) found that about 50\% of the \HI\ in the disk
is ``cold neutral medium'' (CNM), with spin temperatures ranging from 30\,K to
several 100\,K, with a median value of 70\,K. The other half of the \HI\ is
``warm neutral medium'' (WNM) with $T_s$$>$500\,K. Our value of 135\,K for the
CNM is an average. We use the different possible values for $T_s$ to derive a
systematic error (see below). For column densities below \dex{20}\,\cmm2, the
differences between assuming an optically thin cloud and $T_s$=70\,K or
$T_s$=135\,K are $<$1.5\%. Only above column densities of about
5\tdex{20}\,\cmm2\ (log\,$N$(\HI)$>$20.7) does the difference become $>$5\%. In
our sample this only happens for sightlines toward which we compare 21-cm column
densities derived with different radio telescopes, but not for sightlines where
we measure \Lya.
\par The total error on the column density was estimated by adding in quadrature
four sources of error, one statistical and three systematic. First, for each
spectrum the rms in the brightness temperature was calculated and converted to a
corresponding error in column density for the velocity interval that the column
density was integrated over. Second, a systematic error of 1\tdex{18}\,\cmm2\
due to baseline placement was assumed for LAB and GBT data. For the older Green
Bank spectra this was set to 2\tdex{18}\,\cmm2. In one case (3C\,273.0) strong
continuum emission in the beam results in a noticeably worse instrumental
baseline. Therefore, we set the baseline error to 5\tdex{18}\,\cmm2. Similar
effects cause the baseline error for NGC\,985 to be set to 2\tdex{18}\,\cmm2.
The third source of error is the stray radiation correction. Blagrave et al.\
(2010) and Boothroyd et al. (2011, in preparation) study the stray radiation
effects for the GBT in detail and estimate that the stray radiation correction
adds another 4\,K\,\kms\ of uncertainty to an \HI\ profile, equivalent to
7\tdex{18}\,\cmm2\ in \NHI. However, this uncertainty only applies to the
low-velocity emission near 0~\kms. Profile components in the \HI\ spectrum that
have velocities above about 50~\kms\ are not affected as strongly, as there is
much less bright emission elsewhere in the sky at those velocities. Therefore,
for intermediate-velocity components ($v$$\sim$50\,\kms) we used a systematic
error associated with the stray radiation correction of 1\tdex{18}\,\cmm2, while
for high-velocity components ($v$$>$90~\kms) we set this error to zero. The
final source of error is the assumed value for the spin temperature. We
calculated $N$(\HI) using the formula above with $T_s$ set to 70\,K, with
$T_s$=135\,K, and using the optically-thin assumption (reducing the $T$ terms to
$T_B$), and we took the rms variation between those three values as the
corresponding error. The combined error is at most 0.04 dex (10\%) for
sightlines with log\,$N$(\HI)$<$20.8, the highest value we find for a direction
where \Lya\ is measured, but for the great majority of sightlines the combined
systematic error is $<$5\% and usually it is $<$2\%.
\par Finally, we calculated the brightness-temperature weighted average velocity
of the profile, which is used to center the absorption line model (see below).
The velocity intervals were defined by visually determining the extent of the
emission in each of the 21-cm spectra toward a target. In most cases the chosen
interval was then set to be the same at each angular resolution, but in a few
cases small adaptations were necessary. To compare the 21-cm emission to the
\Lya\ absorption, the full extent of the \HI\ emission was used, but high- and
intermediate-velocity components were measured separately when comparing column
densities between different 21-cm observations.

\subsection{GBT data}

\par Of the 59 AGNs with UV data, 35 were newly observed during 2008 and 2009,
using the Green Bank Telescope (GBT) at 0.8~\kms\ spectral resolution and
9\farcm1 angular resolution.
\par The calibration of the GBT data took advantage of a recent, extensive
investigation into the all-sky response of the telescope at 21\,cm, made by a
group which includes one of us (FJL). The detailed results will be given in
Boothroyd et al.\ (2011, in preparation) and are summarized in Blagrave et al.\
(2010). Because the calibration is critical to the \Lya\ vs 21\,cm comparison,
we give a short summary here. Note that the GBT 21-cm and UV calibrations are
completely independent: information from the \Lya\ measurements was never used
in the GBT calibration.
\par The GBT is unique among large single dishes in that it has a clear aperture
and thus does not have the sidelobes caused by scattering from the feed support
legs, subreflector, and other blockage in the aperture (Prestage et al.\ 2009).
The only significant sidelobes arise from spillover past the primary and
secondary reflectors. Thus the telescope response can, to a significant degree,
be derived from theoretical calculations. Boothroyd et al.\ (2011, in
preparation) used calculations to determine the telescope response near the main
beam, and used observations of the Sun to determine the spillover sidelobes. The
aperture efficiency was derived from an electromagnetic analysis of the
telescope which incorporated the detailed telescope geometry and measured
illumination pattern of the 21\,cm receiver. The antenna temperature scale was
then established using this efficiency and observations of the radio source
3C\,286 whose absolute flux was taken from Ott et al.\ (1994). There is
agreement between this antenna temperature scale and one derived from laboratory
measurements of the receiver noise diodes to within 2.4\%. The same calculations
established the main beam efficiency as 0.88. Measurements of the moon (which
fills the GBT beam) were made to check the accuracy of the main beam efficiency
and of point-sources to check the calculated main beam shape. In both cases the
observations agreed with the calculations to within a few percent.
\par The forward and rear spillover lobes were measured by mapping large areas
around the Sun. Given knowledge of the amplitude and location of these 
sidelobes, the Leiden-Dwingeloo 21\,cm \HI\ survey (Hartmann \& Burton 1997) was
then used to estimate the ``stray'' component of 21\,cm emission in the GBT data
which was then removed from the observations. The GBT has very low near
sidelobes (more than 30\,dB below the main beam gain) so no correction is made
for stray radiation arising within one degree of the main beam. The derived
main-beam brightness temperatures were corrected for atmospheric extinction
assuming a zenith opacity of 0.008\%.
\par The GBT 21-cm spectra were reduced and calibrated without reference to
standard \HI\ directions because we felt that for the GBT it would be more
accurate to derive the calibration from fundamental flux density references and
a detailed understanding of the antenna. In addition, the \HI\ brightness
temperature toward the standard directions is known to vary significantly with
the angular resolution of the antenna (Kalberla et al.\ 1982). Nonetheless, the
standard direction S8 was observed four times on two consecutive days during our
experiment, and examination of these data, which were calibrated and corrected
for stray radiation identically to the AGN directions, gives useful information.
The GBT values for both the peak line brightness temperature and the line
integral of S8 are in excellent agreement with those given in Kalberla et al.\
(1982), being in the ratio 0.994$\pm$0.004 and 0.991$\pm$0.001 for the peak and
integral, respectively. The uncertainties are 1$\sigma$ about the mean of the
four observations. Kalberla et al (1982) concluded that when stray radiation is
taken into account the principal \HI\ calibration observations (Penzias et al
1970; Wrixon and Heiles 1972; Williams 1973) are in agreement to within about
3\%. Thus our data reduction procedure appear to produce GBT \HI\ spectra with
an accurate brightness temperature scale, one that is consistent with other
calibrations.
\par Each source was measured by the GBT at least twice, and several as many as
four times. This allows us to estimate an error term based on the
reproducibility of the results. The short-term reproducibility is derived from
sequential scans; here the main sources of error should be noise and
instrumental baseline changes. We find median short-term differences in the line
integral of 0.4\%, corresponding to median fluctuations in \NHI\ of
1.1\tdex{18}\,\cmm2. This is what would be expected from noise or baseline
fluctuations already included in our error analysis.
\par Long term differences in spectra test not only baseline and gain  
stability, but the accuracy of the stray radiation correction. The median
long-term uncertainty in the line integral is 2.3\%, equivalent to
\NHI=1.9\tdex{18}\,\cmm2. These differences lie well within that expected from
uncertainty in the GBT stray radiation correction.
\par All of the tests that we have been able to make indicate that our
fundamental calibration of the GBT \HI\ spectra is accurate, and that our error
estimates give a faithful representation of the uncertainties. At no time during
this experiment, or during others that have measured many \dex4\ \HI\ spectra
with the GBT (e.g.\ Lockman et al.\ 2008), have 10\% fluctuations in the GBT
\HI\ intensity scale been observed like those reported by Robishaw \& Heiles
(2009).

\section{Systematic effects in the \HI\ data}
\par In this section we discuss the comparison between the column densities
measured using \Lya\ absorption and 21-cm emission. A straightforward comparison
between the values of \NHI\ measured from 21-cm emission with those measured
using various radio telescopes reveals significant differences between the two.
In particular, for our set of low-column density directions, the values found
with the Green Bank 140-ft and from the LAB survey are on average about 10\%
larger than those derived from \Lya. On the other hand, no significant
difference is found when comparing GBT and \Lya\ column densities. Below we will
show why we think that these differences arise from the presence of a broad,
spurious component in the LAB and 140-ft spectra. The amplitude and FWHM of this
component are 0.048~K and 167~\kms\ for LAB spectra, 0.023~K and 134~\kms\ for
140-ft data with 2~\kms\ channels, and 0.12~K and 70~\kms\ for 140-ft data with
1~\kms\ channels, corresponding to column densities of 15.3\tdex{18},
5.9\tdex{18} and 15.8\tdex{18}\,\cmm2.
\par We discovered this after our referee pointed us to a website commonly used
to obtain column densities from the LAB data
(http://heasarc.gsfc.nasa.gov/cgi-bin/Tools/w3nh/w3nh.pl). Although this helped
us to figure out the problems, we want to point out that we have concluded that
this website gives incorrect answers. In particular, it gives column densities
that are derived after interpolating between the pixels of a gridded all-sky map
smoothed (by default) to a one degree beam. Thus, a varying number of original
LAB spectra is used to derive a column density. When using half a degree as the
smoothing radius there are directions where the website says that there are no
data on positions where an actual observation was done. Furthermore, in most
cases the column density that it gives for a position where an actual spectrum
was taken differs from the column density derived from that actual spectrum.
\par However, comparing our numbers with those given by this website led us to
the realization that integrating LAB spectra from $-$400 to 400~\kms\ gives
different results than integrating over the line core, i.e., the velocity range
where emission appears present. We had done the latter, in order to increase the
S/N ratio by avoiding integrating over a 300~\kms\ window containing only noise,
as well as to be able to separate out high-velocity emission. For the LAB
spectra, the two different integrals differ by a constant offset of
$\sim$6\tdex{18}\,\cmm2, as shown in Fig.~\Fcompare. Since the highest \HI\
column density in the set of sightlines with \Lya\ data is only about
7\tdex{20}\,\cmm2, we added 90 sightlines with higher column density to make
this comparison. These 90 sightlines were chosen at latitudes $b$=0\deg\ to
$b$=30\deg, at longitudes 170\deg, 180\deg\ and 190\deg. This extends the
comparison to \NHI$\sim$7\tdex{21}\,\cmm2.
\par The origin of the offset seen in Fig.~\Fcompare\ becomes clear if we
examine the parts of the 21-cm spectra outside the bright line cores.
Figure~\Fnosignal\ shows the resulting averages of the non-signal regions of
each spectrum. That is, for each direction we selected the channels outside the
velocity limits given in Table~\Tres. These limits are based on visually
inspecting the spectra and selecting the velocity range where emission appears
present. We did this for the GBT, 140-ft and LAB spectra in the same set of
directions. Panel (a) gives the residual spectrum for the combined set of all
LAB directions (i.e., the those with \Lya, GBT and 140-ft, as well as the higher
column densities directions used to make Fig.~\Fcompare). Panels (d), (e) and
(f) show the residuals for the two kinds of 140-ft spectra (with 2~\kms\ and
1~\kms\ channel spacing), and for the GBT spectra. Further, in panels (b) and
(c) we show the residual LAB spectra for the directions corresponding to the
140-ft and the GBT sightlines. Gaussians were fit to the resulting features in
order to determine the total emission in this broad component. Note that the GBT
data show no significant broad component, although we show the formal fit.

\begin{figure}\plotfiddle{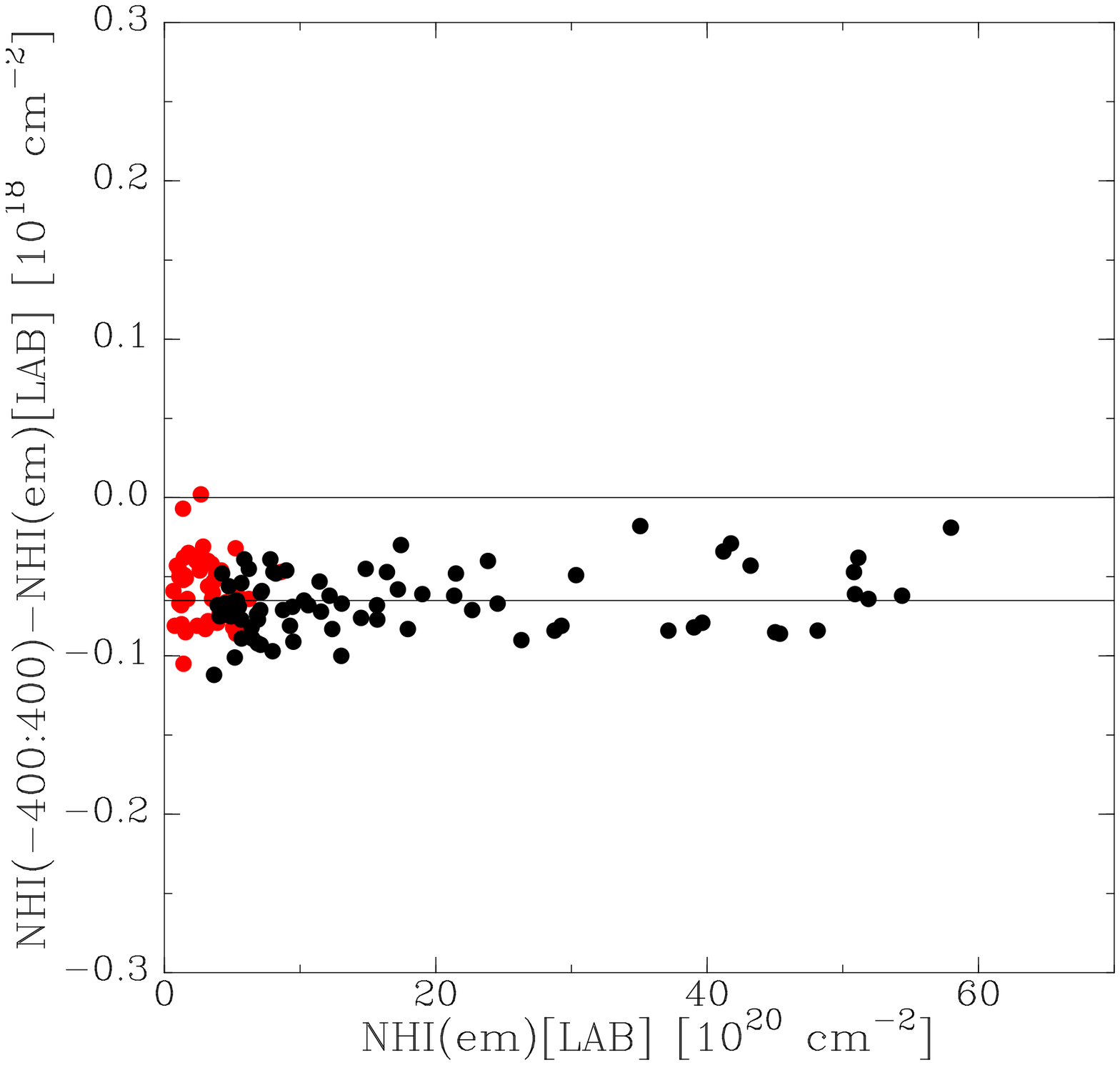}{0in}{0}{300}{300}{0}{0}\figurenum{\Fcompare}
\caption{Comparison between column densities derived from LAB spectra when
integrating from $-$400 to 400~\kms\ (\NHI($-$400:400)) to values derived when
integrating only over the region where emission is visible (\NHI(em)). Red
points are for the directions toward the AGNs for which we analyzed \Lya\
spectra. Black points are for higher-column density directions along three
strips at $l$=170\deg, 180\deg\ and 190\deg, $b$=0\deg\ to 30\deg, in steps of
1\deg. A least squares-fit to these differences gives a slope of zero and an
offset of 0.65\tdex{18}\,\cmm2.}
\end{figure}

\begin{figure}\plotfiddle{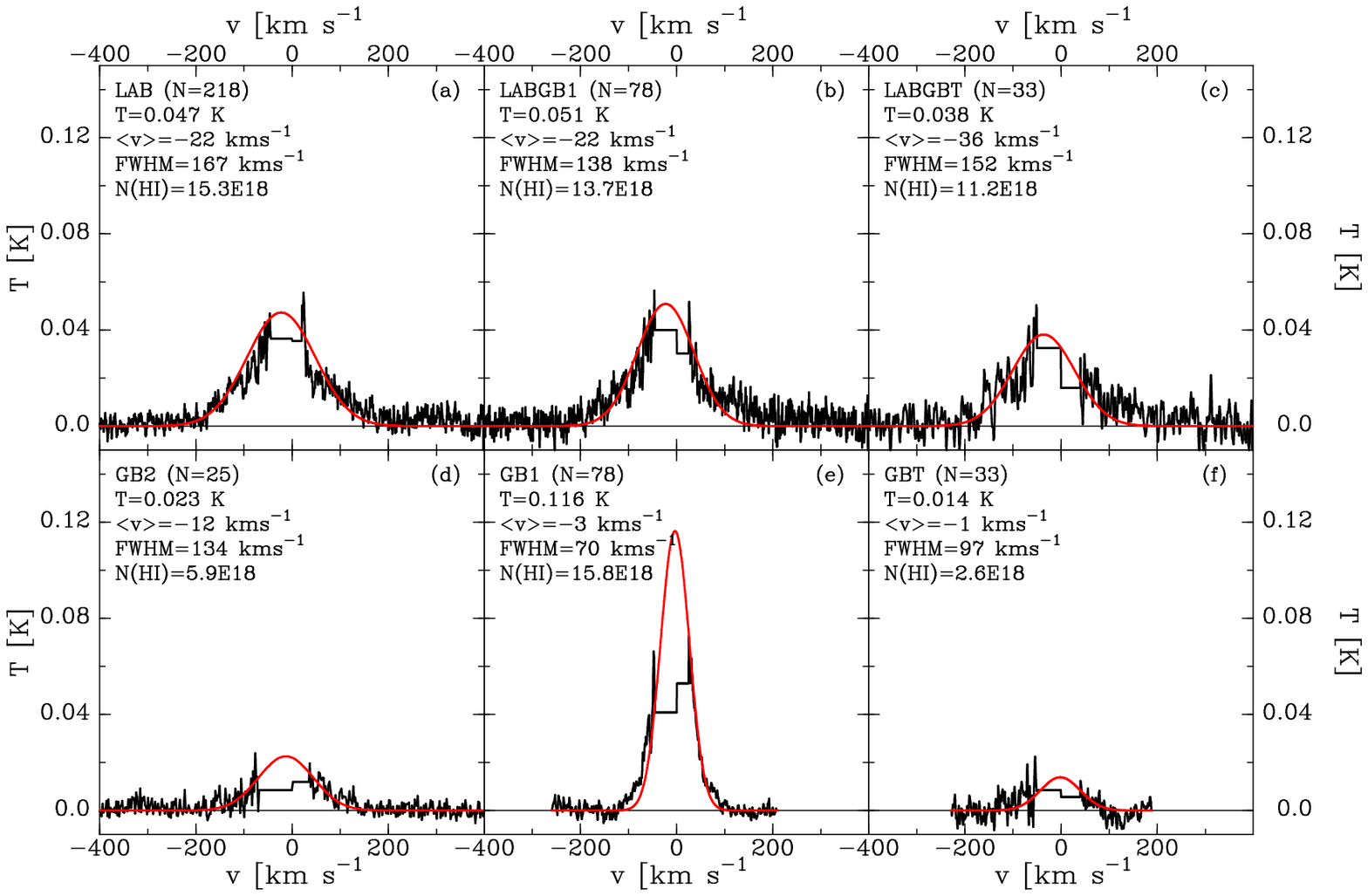}{0in}{0}{440}{300}{0}{0}\figurenum{\Fnosignal}
\caption{Black lines show the residual spectra for LAB, Green Bank 140-ft and
GBT data, created by cutting out the region in each individual spectrum where
\HI\ emission is visually present. Panel (a) is for LAB spectra in the combined
set of directions formed by our GBT, 140-ft spectra and the two strips near
$l$=180\deg\ used to create Fig.~\Fcompare. Panels (d) and (e) are for 140-ft
spectra, with panel (d) for spectra having 2~\kms\ channel spacing and panel (e)
for spectra with 1~\kms\ channels. Panel (b) shows the LAB residual for the
140-ft directions. Panel (f) is the residual for the GBT spectra, while panel
(c) is the LAB residual for the GBT directions. The red lines are gaussian fits
to the residuals, with the parameters of the gaussian given in the label near
the top of each panel. The label also gives the number of spectra used to make
these residuals (N=\#), and the integrated column density corresponding to the
gaussian.}
\end{figure}

\par The broad residual component in the LAB spectra has a FWHM of 167~\kms\
(dispersion 71~\kms) and a column density of 1.5\tdex{19}\,\cmm2. We note that a
similar broad underlying component in the LAB spectra was originally found by
Kalberla et al.\ (1998), who measured a component with dispersion 60~\kms\ (FWHM
140~\kms) and total column density 1.4\tdex{19}\cmm2. Kalberla et al.\ (1998)
interpreted this as evidence for a halo component in the Galactic \HI, with
scaleheight 4.4~kpc. However, we will now show that the overwhelming
preponderance of evidence points to the conclusion that this broad underlying
component is an artifact. This is based on the following four arguments.
\par {\it (1) There is no residual component in GBT spectra.} Since
theoretically the GBT has very stable baselines, and is less affected by stray
radiation than the LAB dataset, we weigh the absence of a residual in the GBT
spectra as an argument in favor of the conclusion that the residual seen in the
LAB spectra is an artifact.
\par {\it (2) The residual emission seen in LAB and 140-ft spectra differs.}
Where the LAB spectra produce a residual feature that is a gaussian with
$T$=0.048~K and FWHM=167~\kms, the 140-ft data with 2~\kms\ channels give a
gaussian with $T$=0.023~K, FWHM=134~\kms. Notably, 140-ft with 1~\kms\ channel
spacing give $T$=0.12~K, FWHM=70~\kms. Thus, the width of the gaussian differs
for the two sets of similarly-calibrated but differently-setup 140-ft data, even
though the same velocity ranges in the same spectra were used to create these
residuals. Further, both residuals differ from that in the LAB data. This again
suggests that the residual emission is an artifact. Still, it is possible that
the calibration of the LAB spectra was better and the broad component is only
properly picked up in that dataset. On the other hand, the difference in
residual between the two kinds of 140-ft data suggests that it may be caused by
a (small) baseline fitting error.
\par {\it (3) The broad component produces no metal-line absorption.} {\it If}
there is a broad \HI\ component in the Galaxy, this should result in a
detectable optical depth for many ionic absorption lines in the ultraviolet. For
instance, for solar metallicity gas with typical dust content, a component with
FWHM 170~\kms\ will produce a peak optical depth of $>2$ in the \CII$\
\lambda\lambda$1334.532, 1036.337, \OI\ $\lambda$1302.169 and \SiII\
$\lambda$1260.422 lines. This would yield a dark line out to $\pm$150~\kms\ in
{\it every} sightline. As can be seen from the spectra toward 100 AGNs presented
by Wakker et al.\ (2003), this is clearly not observed. A typical example for a
direction without any known 21-cm emitting high-velocity gas is shown in
Fig.~\Fexample. In fact, in all directions the extent of the strong metal lines
is rather similar to the visual extent of the \HI\ emission. Figure~\Fexample\
also includes the absorption profile that would be associated with the
high-dispersion gas if it had 1/10th solar metallicity. This could be the case
if the high-dispersion gas consisted of infalling gas. However, in that case
there would still be clear extended wings, which are also not observed. Thus, if
this high-dispersion component were Galactic gas, there would be unmistakeable
evidence for it in these spectra, {\it unless} its metallicity is much less than
0.1 times solar.

\begin{figure}\plotfiddle{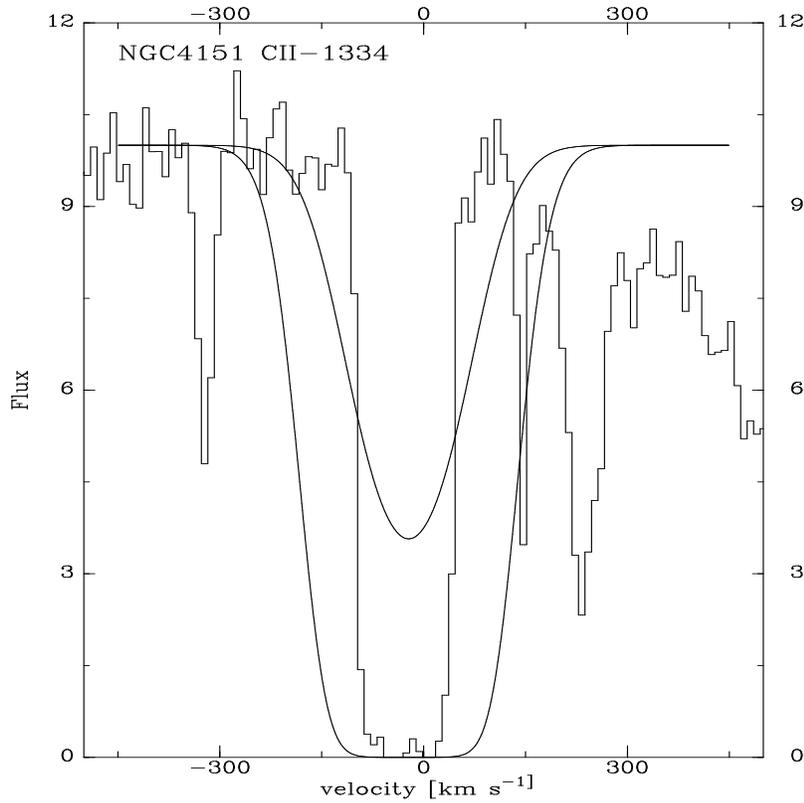}{0in}{0}{300}{300}{0}{0}\figurenum{\Fexample}
\caption{A typical example of the observed \CII\ $\lambda$1334.532 absorption
line compared to the expected absorption if there were high-dispersion material
in the \HI\ profile with FWHM 167~\kms\ and $T_B$=0.048~K. Two smooth
theoretical curves are shown on top of the data, one for solar metallicity
(going down to having no flux in the center) and one for 1/10th solar
metallicity. In neither case are the broad wings expected from a high
velocity-dispersion interstellar component seen in the actual \CII\ line.}
\end{figure}

\par {\it (4) The broad component can not originate as an average of a
population of small clouds.} The only way in which a broad component might be
absent in absorption line spectra, is if it consists of many small clumps, which
are missed in all of the 100 or so observed AGN sightlines. Since 100 sightlines
were sampled, to have a 50--50 probability of missing the small clouds in {\it
all} sightlines requires that they have an area covering fraction of $<$0.7\%,
since 0.993$^{100}$=0.5. To have one such cloud in every 0\fdg5 beam thus
requires clouds with diameters that are less than 3\arcmin. As the observed
brightness temperature at 36\arcmin\ resolution is 0.048~K, this implies an
actual brightness temperature of 7~K, which for clouds with linewidth 15~\kms\
would imply a colum density of 2\tdex{20}. When observed with a 10\arcmin\ beam,
these clouds would have apparent brightness temperatures of 0.6~K and apparent
column densities of \dex{19}\,\cmm2. The population should have a cloud-to-cloud
dispersion of 70~\kms, and be widespread at low {\it and} high latitudes.
\par A population of small clouds has indeed been discovered (Lockman 2002; Ford
et al.\ 2008, 2010). These studies find 400 and 255 small clouds in two 720
square degree regions (0.45 per square degree), with the clouds having typical
peak brightness temperature 0.5~K, typical velocity width 13~\kms, and typical
size 20\arcmin\ (i.e.\ many are unresolved). However, although the brightness
temperature of these clouds is similar to what is required, they are much too
large, the population strongly is confined to low latitudes, and the
cloud-to-cloud dispersion is only 16~\kms. Thus, these clouds do {\it not} fit
the requirements, and the observations show that a population of clouds with the
required characteristics to mimic the broad component in the LAB spectra does
{\it not} exist. 
\par Combining points (1) through (4) leads to the conclusion that the residual
emission in the LAB spectra is an artifact. The residual could be due to a final
imperfection in the stray radiation correction. It might also be caused by a
small error in the polynomials used to fit the baselines. Such an error would be
difficult to discern since the largest visible deviation is $\sim$0.02~K, or
less than 1/3rd the rms noise in a single spectrum. The fact that the spectra
from different telescopes have different residuals argues for the second option.
The fact that no significant residual is seen in the GBT spectra argues that the
first option might also play a role.
\par Based on these considerations, we decided to correct the LAB and the 140-ft
spectra, subtracting the gaussians shown in Fig.~\Fnosignal. This eliminates the
discrepancies between the column densities derived from \Lya\ and those derived
from 21-cm observations. As we show in the next section, the average ratio
\NHILya/\NHITWcm\ then becomes 0.96--1.00, with a dispersion of $\sim$0.10, not
significantly different from 1. Without this correction, the average ratio is
0.90$\pm$0.10 for the LAB data and the LAB spectra are in tension with the \Lya\
column densities, with the GBT spectra, with the non-detection of metal-line
absorption and with the observed population of small clouds. However, after
applying the corrections, all tension between the many different observations
disappears. In the remainder of the paper, we will thus only use corrected LAB
and 140-ft column densities. We leave the GBT column densities uncorrected
because there is no strong evidence that a correction is needed.
\par We also note the following. In Sect.~\SGBT\ we had said that for the
standard field S8 comparing the column density derived from the GBT data to that
derived from the LAB survey gave a ratio of 0.991. The actual LAB column density
toward S8 is 1.805\tdex{21}\,\cmm2. Subtracting the spurious
1.5\tdex{19}\,\cmm2\ from this gives a column density of 1.790\tdex{21}\,\cmm2.
The ratio of these two values is 0.992. Thus, after correcting the LAB data,
they give exactly the same column density as the GBT data in this direction.

\section{Results}

\subsection{21-cm and UV data}
\par In Fig.~\Fspectra\ we show the 21-cm and UV spectra of all AGN targets
toward which we measured \NHILya. The 21-cm spectrum that is displayed is the
one obtained with the smallest telescope beam. In the panels with the UV
spectra, we show the observed flux as well as the fitted continua and resulting
models. The \HI\ column densities derived from the 21-cm and \Lya\ absorption
are shown for comparison. Table~\Tres\ presents a more detailed summary of the
results, giving the UV column densities, as well as the 21-cm column densities
obtained with different telescopes (see Sect.~\Sobs).
\par The spectra of targets observed with the \STIS-G140M grating sometimes show
a rise on the short-wavelength side of the spectrum, below 1200~\AA. We have
been unable to determine the origin of this spectral slope, but it may be due to
a calibration error in G140M observations. The targets that show this effect
are: Mrk\,1044, Mrk\,1513, PG\,0804+761, PG\,1049$-$005, PG\,1341+258,
PKS\,2005$-$489, RX\,J0100.4$-$5113, RX\,J1830.3+7312, Ton\,S180, and
VII\,Zw118, as can be seen in Fig.~\Fspectra. Where possible, we avoided fitting
the continuum using pixels below 1200~\AA.

\begin{figure}\plotfiddle{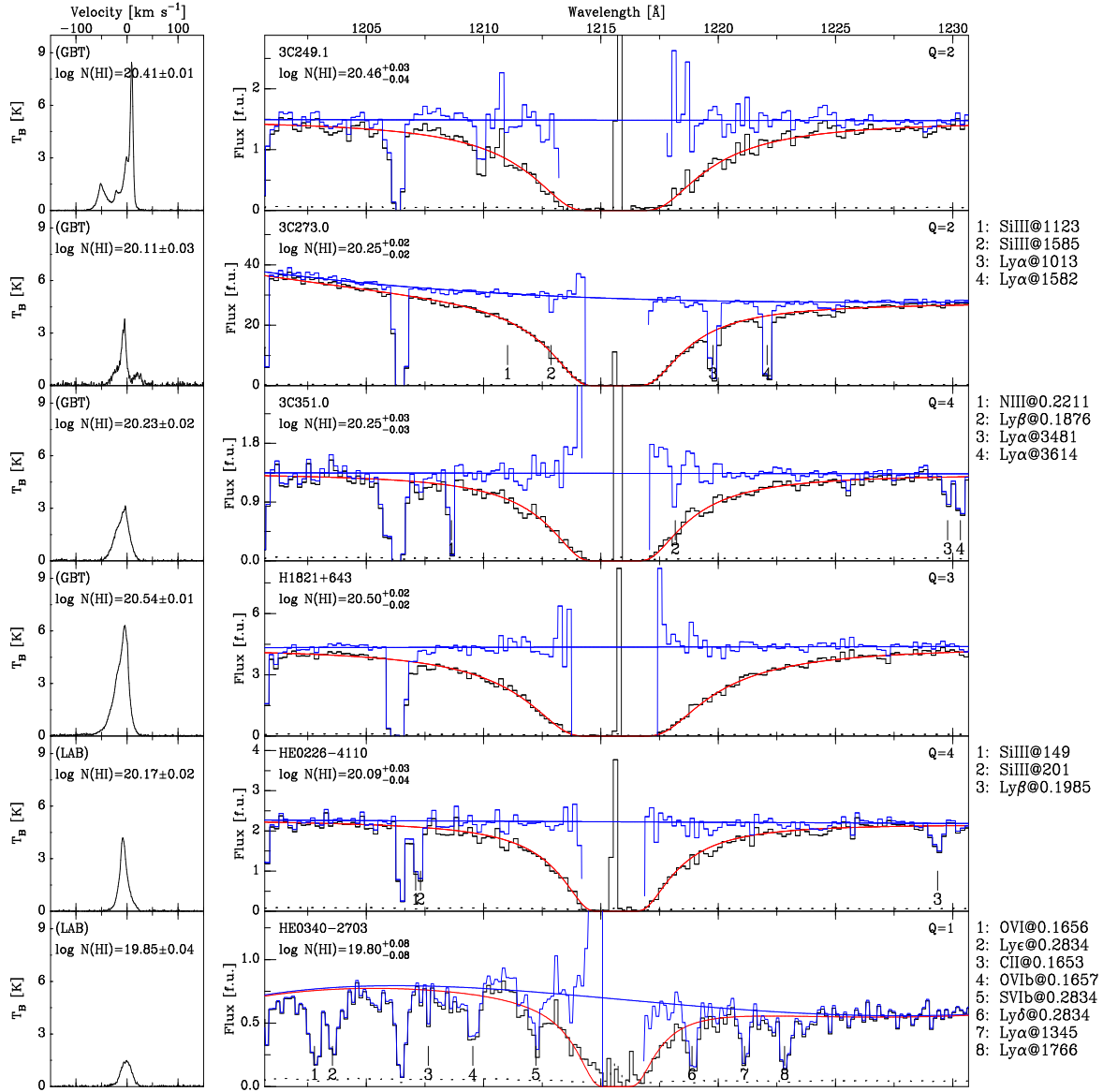}{0in}{0}{440}{440}{10}{0}\figurenum{\Fspectra a}
\caption{
Left: 21-cm brightness temperature versus LSR velocity in either
Leiden-Argentina-Bonn survey, Green Bank 140-ft, or Green Bank Telescope (GBT)
observations, as shown by the labels LAB, GB, and GBT. The available spectrum
with the smallest beam on the sky is shown. The second label line gives
log\,\NHITWcm\ for each sightline. Right: Black histograms give the \Lya\ flux
versus wavelength from \HST\ \STIS-G140M and \STIS-E140M data. The blue
histograms represent the flux after correction for the Galactic \Lya\
absorption. The smooth lines through the blue histograms are the reconstructed
continua, which were fitted through the corrected data; see Sect.~\Sfitmethod\
for a complete description. The solid red lines give the final model fits to the
data. The fits are calculated using the reconstructed continuum and the value of
\NHILya\ that is listed in the label on the left. The quality factor for each
sightline (see Sect.~\Sdataqual) is given in the label on the right. The dashed
line at the bottom of each spectrum represents the error in the observed fluxes.
For selected sightlines IGM or ISM emission or absorption lines are identified
to the right of the spectra with corresponding numbers placed below the spectra
and the heliocentric velocity of the line included in the label.
}\end{figure}
\begin{figure}\plotfiddle{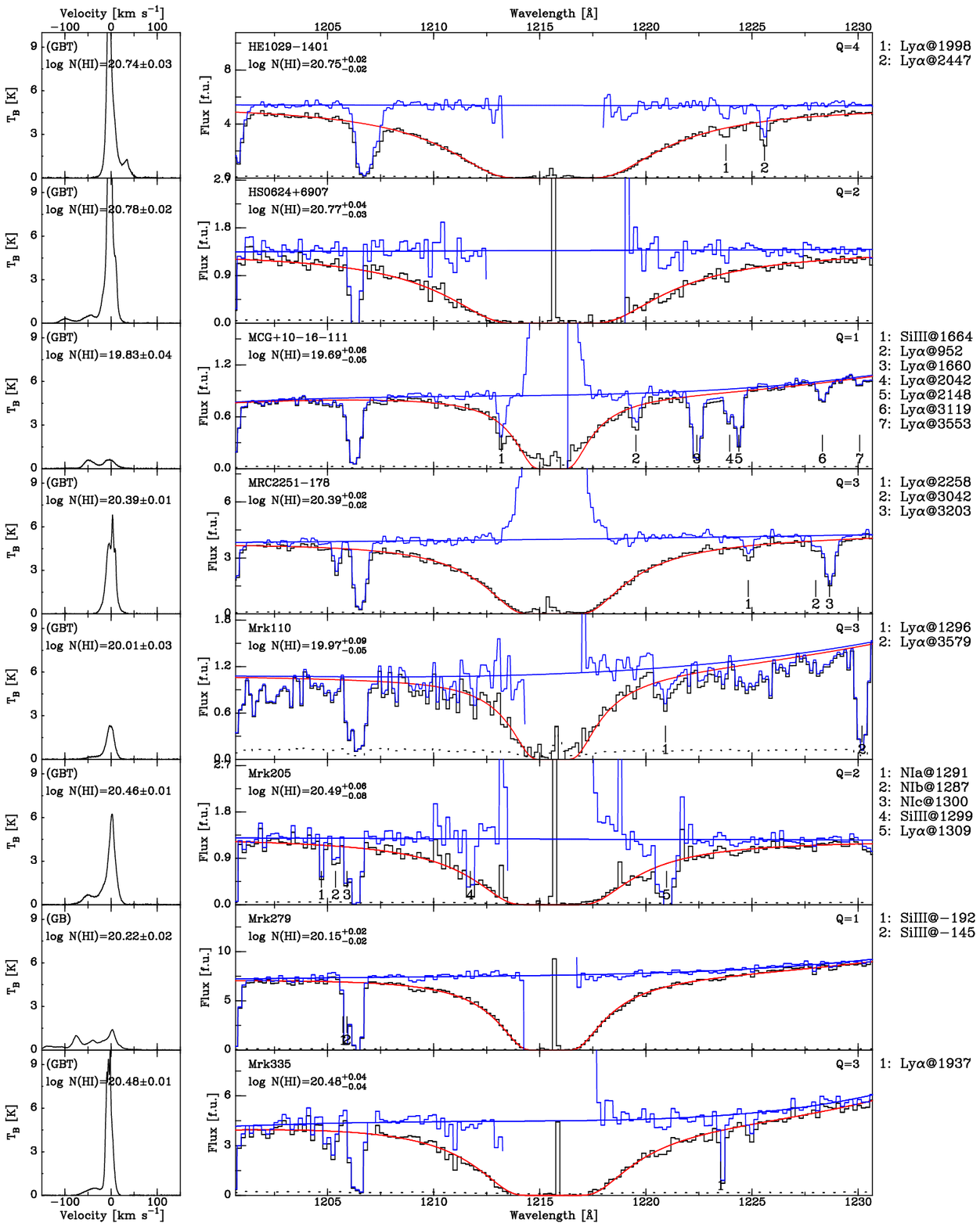}{0in}{0}{440}{440}{10}{0}\figurenum{\Fspectra b}\caption{Continued.}\end{figure}
\begin{figure}\plotfiddle{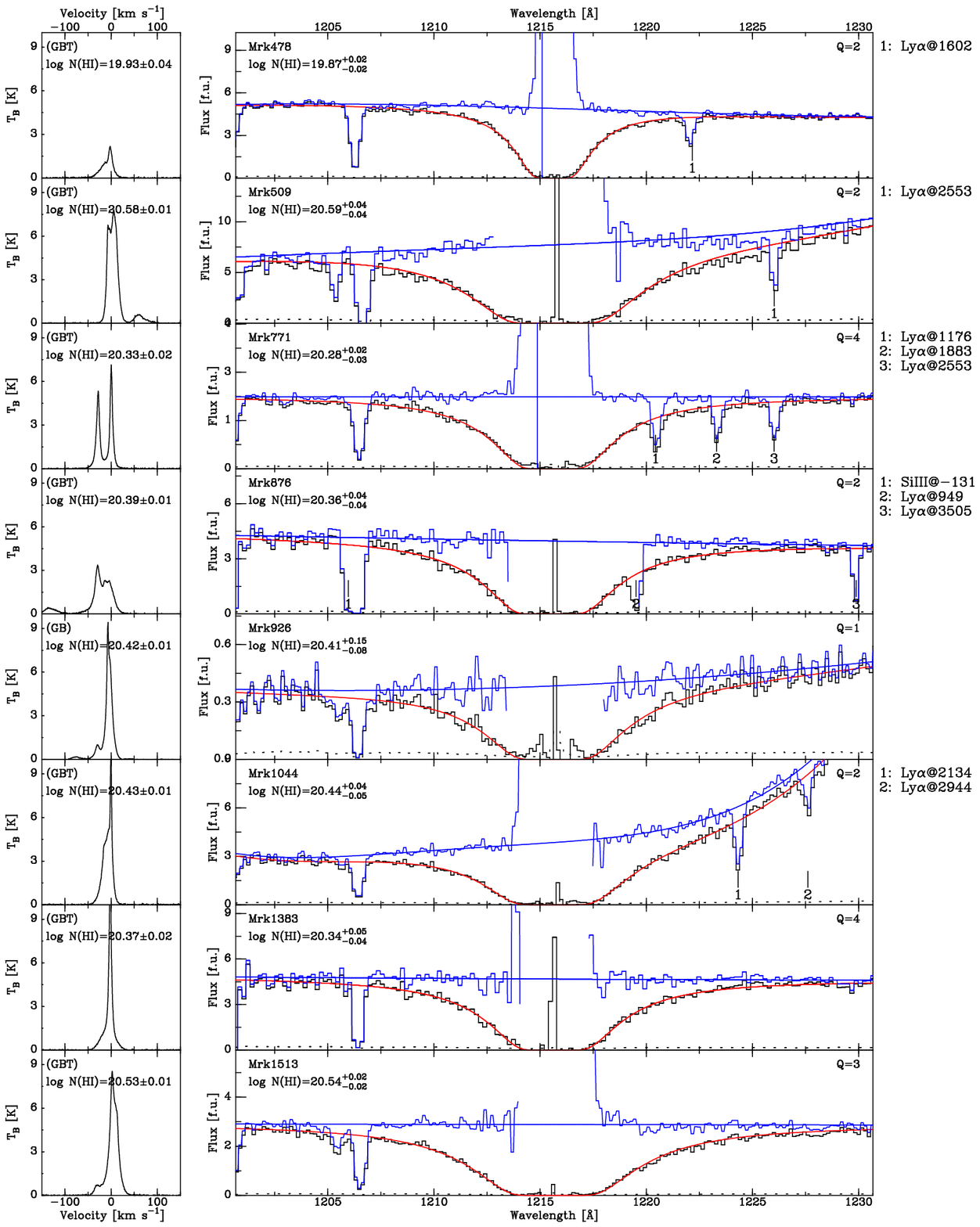}{0in}{0}{440}{440}{10}{0}\figurenum{\Fspectra c}\caption{Continued.}\end{figure}
\begin{figure}\plotfiddle{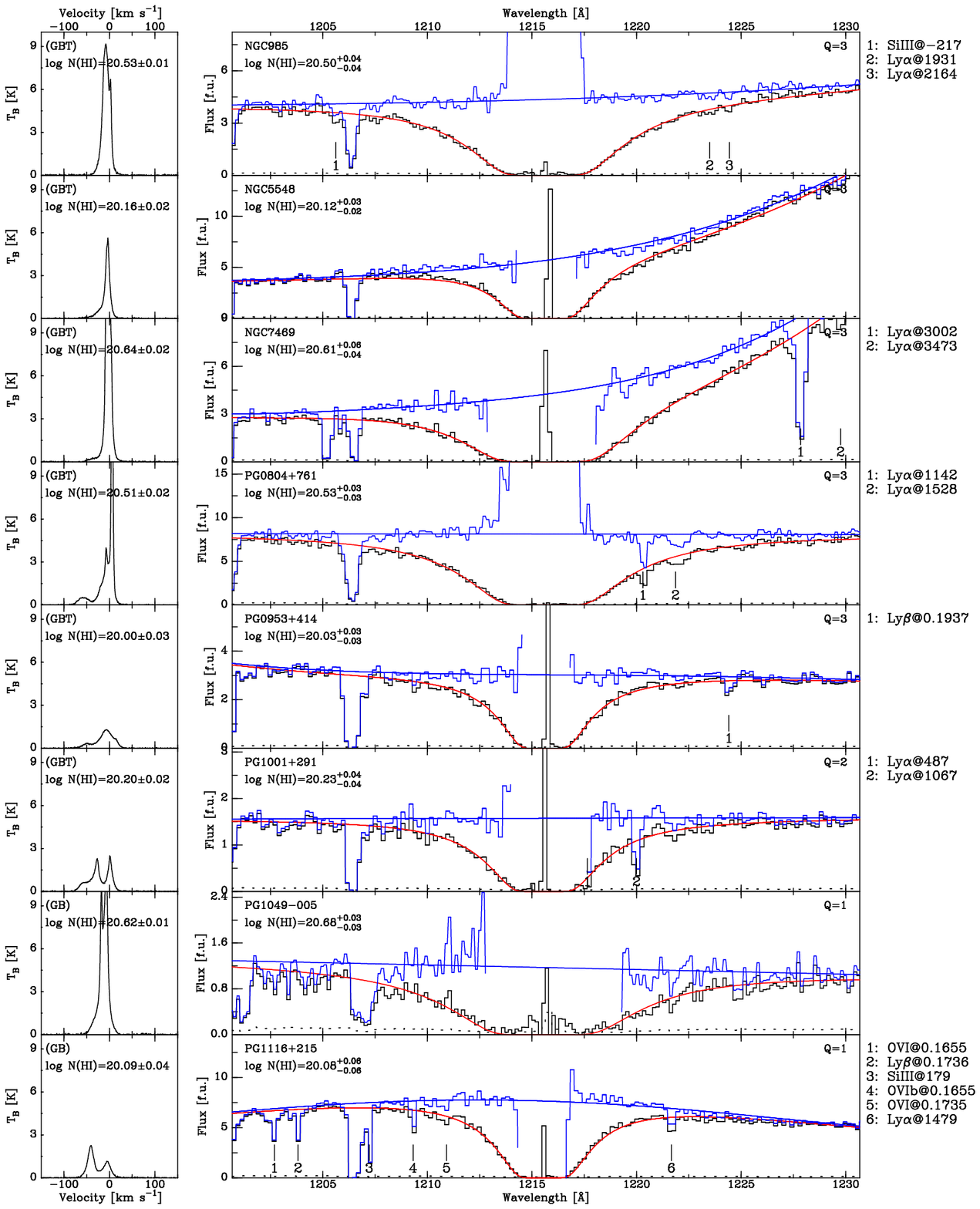}{0in}{0}{440}{440}{10}{0}\figurenum{\Fspectra d}\caption{Continued.}\end{figure}
\begin{figure}\plotfiddle{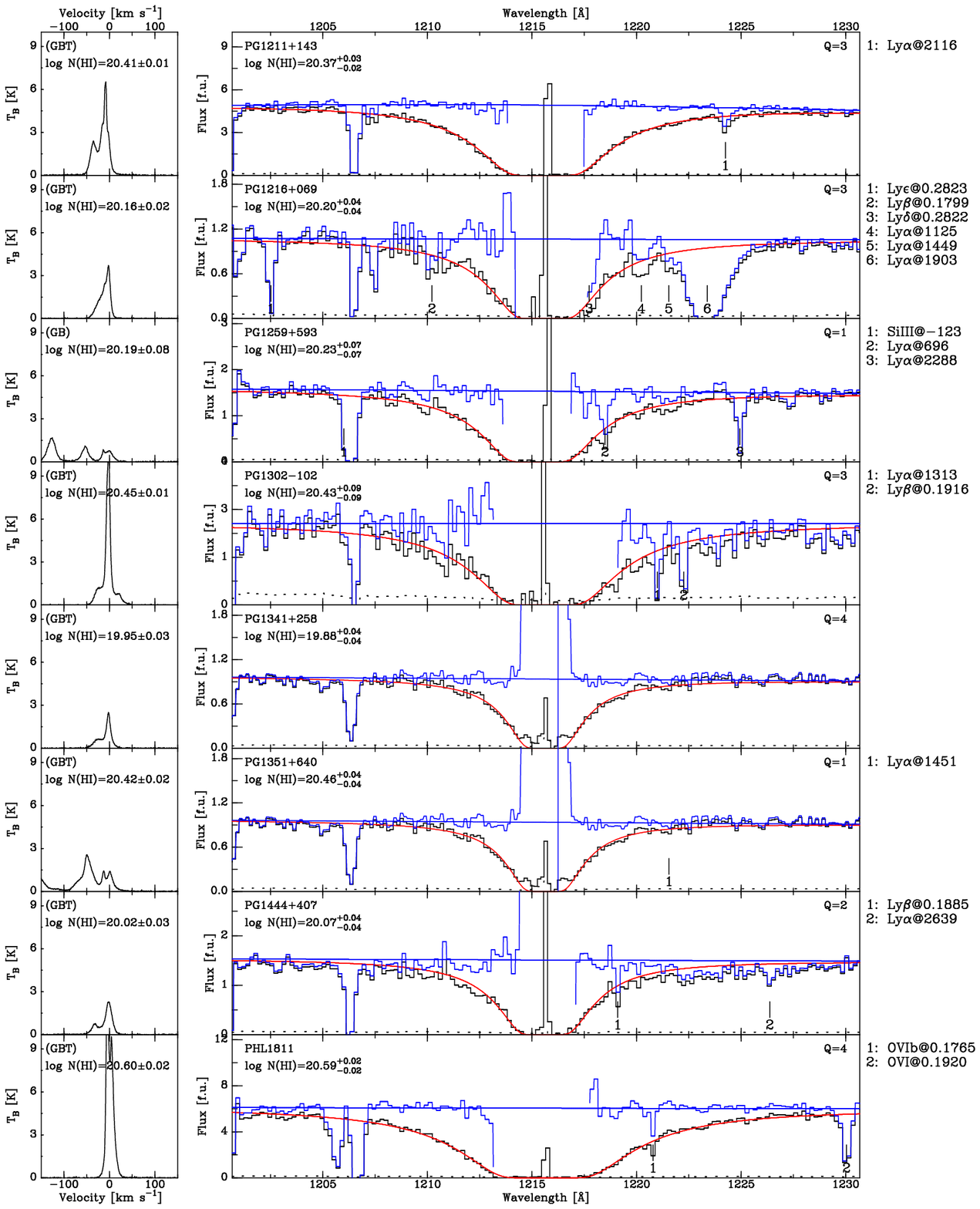}{0in}{0}{440}{440}{10}{0}\figurenum{\Fspectra e}\caption{Continued.}\end{figure}
\begin{figure}\plotfiddle{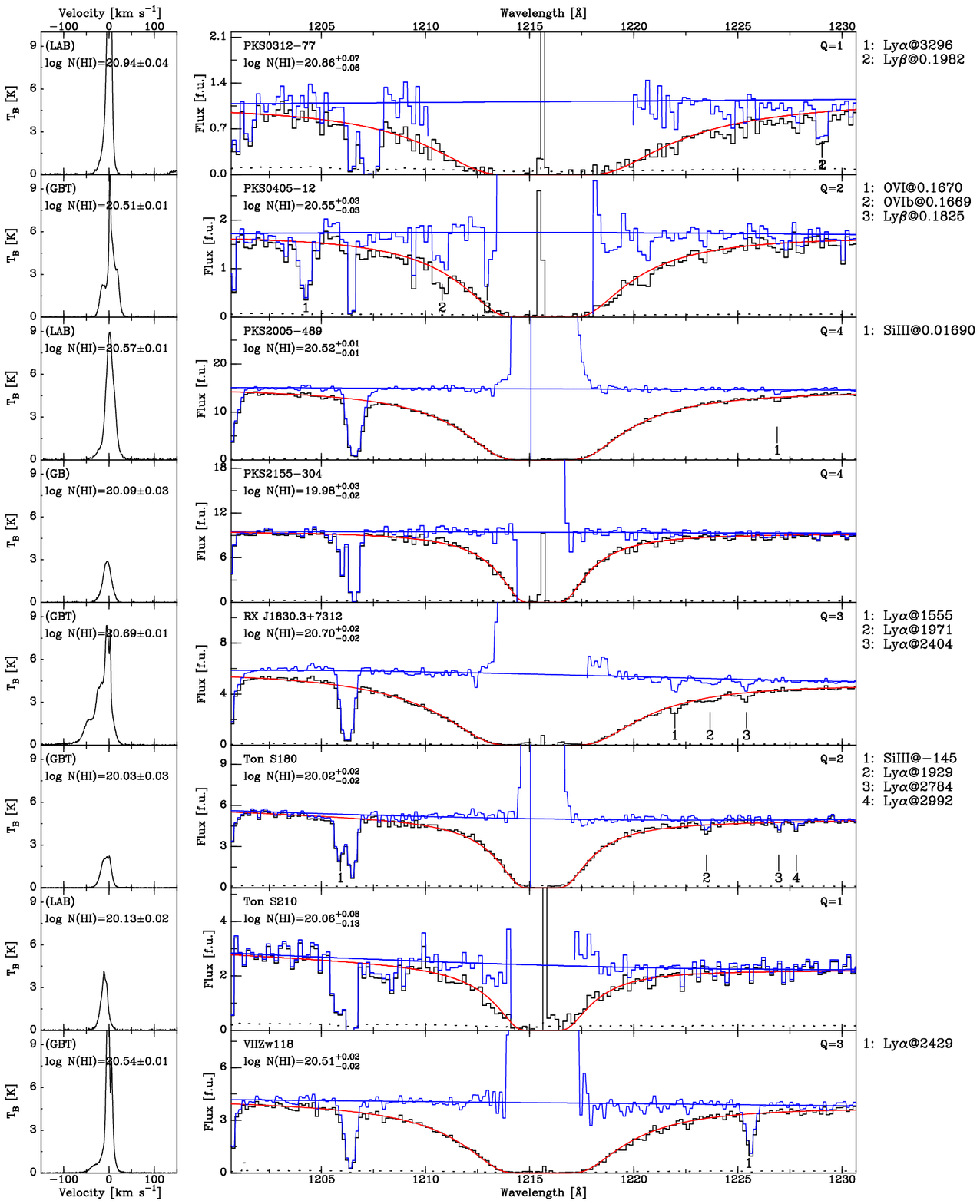}{0in}{0}{440}{330}{10}{0}\figurenum{\Fspectra f}\caption{Continued.}\end{figure}

\def\e{$\pm$}
\def\vmin{$v_{\rm min}$}
\def\vmax{$v_{\rm max}$}
\def\NH{$N$(HI)}
\begin{deluxetable}{lrrccrrrcrcrcr}
\tabletypesize{\scriptsize} \tabcolsep=3pt 
\tablenum{2} \tablewidth{0pt}
\tablecolumns{14}
\tablecaption{Results}
\rotate
\tablehead{%
\colhead{Object}&\colhead{Lon}   &\colhead{Lat}    &\colhead{Qual} &\colhead{log(\NH)} &\colhead{\NH(fixed)}      &\colhead{\vmin} &\colhead{\vmax}  &\colhead{\NH}    &\colhead{$<v>$}   &\colhead{\NH}    &\colhead{$<v>$}   &\colhead{\NH}     &\colhead{$<v>$}  \\
                &\colhead{[\deg]}&\colhead{[\deg]} &\colhead{}     &\colhead{[\cmm2]}  &\colhead{[\dex{18} \cmm2} &\colhead{[\kms]}&\colhead{[\kms]} &\colhead{[\cmm2]}&\colhead{[\kms]}  &\colhead{[\cmm2]}&\colhead{[\kms]}  &\colhead{[\cmm2]} &\colhead{[\kms]} \\
                &                &                 &               &\colhead{(\Lya)}   &\colhead{@\kms]}          &                &                 &\colhead{\ \ \ \ \ \ \ \ \ \ (LAB)}&&\colhead{\ \ \ \ \ \ \ \ \ \ (GB)} &&\colhead{\ \ \ \ \ \ \ \ \ \ (GBT)}&\\
\colhead{(1)}   &\colhead{(2)}   &\colhead{(3)}    &\colhead{(4)}  &\colhead{(5)}      &\colhead{(6)}             &\colhead{(7)}   &\colhead{(8)}    &\colhead{(9)}    &\colhead{(10)}    &\colhead{(11)}   &\colhead{(12)}    &\colhead{(13)}    &\colhead{(14)}   \\
}
\startdata
3C249.1             & 130.39 &  38.55 & 2 &  20.46$_{-0.04}^{+0.03}$  &                &  $-$90~ &  35~~~~ &  20.400\e0.013  &  $-$11~~~ &   20.431\e0.013  &  $-$11~~~ &  20.414\e0.013  &  $-$9~~~\\
3C273.0             & 289.95 &  64.36 & 2 &  20.25$_{-0.02}^{+0.02}$  &                &  $-$55~ &  50~~~~ &  20.169\e0.026  &   $-$8~~~ &   20.184\e0.021  &   $-$4~~~ &  20.108\e0.032  &  $-$4~~~\\
3C351.0             &  90.08 &  36.38 & 4 &  20.25$_{-0.03}^{+0.03}$  &                & $-$170~ &  40~~~~ &  20.220\e0.019  &  $-$10~~~ &                   &         &  20.226\e0.018  & $-$11~~~\\
H1821+643           &  94.00 &  27.42 & 3 &  20.50$_{-0.02}^{+0.02}$  &                & $-$150~ &  40~~~~ &  20.528\e0.012  &  $-$12~~~ &                   &         &  20.536\e0.011  & $-$14~~~\\
HE0226$-$4110$^a$     & 253.94 & $-$65.77 & 4 &  20.09$_{-0.04}^{+0.03}$  &                &  $-$75~ &  60~~~~ &  20.171\e0.021  &   $-$6~~~ &                   &         &                   &       \\
HE0340$-$2703         & 222.68 & $-$52.12 & 1 &  19.80$_{-0.08}^{+0.08}$  &                &  $-$50~ &  30~~~~ &  19.848\e0.042  &   $-$2~~~ &                   &         &                   &       \\
HE1029$-$1401         & 259.33 &  36.52 & 4 &  20.75$_{-0.02}^{+0.02}$  &                &  $-$45~ & 100~~~~ &  20.740\e0.030  &         &   20.757\e0.028  &    1~~~ &  20.736\e0.029  &       \\
HS0624+6907         & 145.71 &  23.35 & 2 &  20.77$_{-0.03}^{+0.04}$  &                & $-$125~ &  50~~~~ &  20.791\e0.025  &   $-$8~~~ &                   &         &  20.781\e0.024  &  $-$7~~~\\
MCG+10$-$16$-$111       & 144.21 &  55.08 & 1 &  19.84$_{-0.04}^{+0.04}$  &                & $-$120~ &  50~~~~ &  19.736\e0.056  &  $-$20~~~ &                   &         &  19.830\e0.043  & $-$24~~~\\
                    &   0.00 &   0.00 & 0 &  19.69$_{-0.05}^{+0.06}$  &  20@$-$47        &  $-$25~ &  50~~~~ &  19.498\e0.093  &   $-$1~~~ &                   &         &  19.552\e0.079  &  $-$1~~~\\
                    &   0.00 &   0.00 & 0 &  19.62$_{-0.06}^{+0.06}$  &  26@$-$47\\
                    &   0.00 &   0.00 & 0 &  19.53$_{-0.08}^{+0.09}$  &  32@$-$47\\
                    &   0.00 &   0.00 & 0 &  19.53$_{-0.05}^{+0.08}$  &  31@$-$1         & $-$120~ & $-$25~~~~ &  19.361\e0.123  &  $-$45~~~ &                   &         &  19.505\e0.087  & $-$49~~~\\
                    &   0.00 &   0.00 & 0 &  19.45$_{-0.07}^{+0.07}$  &  39@$-$1\\
                    &   0.00 &   0.00 & 0 &  19.34$_{-0.09}^{+0.09}$  &  47@$-$1\\
MRC2251$-$178         &  46.20 & $-$61.33 & 3 &  20.39$_{-0.02}^{+0.02}$  &                &  $-$50~ &  50~~~~ &  20.374\e0.015  &   $-$2~~~ &   20.415\e0.016  &         &  20.390\e0.014  &       \\
Mrk110              & 165.01 &  44.36 & 3 &  19.97$_{-0.05}^{+0.09}$  &                &  $-$85~ &  60~~~~ &  20.063\e0.027  &   $-$6~~~ &   20.056\e0.027  &   $-$2~~~ &  20.010\e0.029  &  $-$6~~~\\
Mrk205              & 125.45 &  41.67 & 2 &  20.49$_{-0.08}^{+0.06}$  &  0@$-$199        & $-$225~ &  50~~~~ &  20.436\e0.012  &  $-$24~~~ &                   &         &  20.463\e0.012  & $-$18~~~\\
                    &   0.00 &   0.00 & 0 &  20.48$_{-0.08}^{+0.06}$  &  8@$-$199\\
                    &   0.00 &   0.00 & 0 &  20.47$_{-0.08}^{+0.06}$  &  16@$-$199\\
                    &   0.00 &   0.00 & 0 &  20.46$_{-0.09}^{+0.06}$  &  24@$-$199\\
Mrk279              & 115.04 &  46.86 & 1 &  20.15$_{-0.02}^{+0.02}$  &                & $-$175~ &  40~~~~ &  20.128\e0.023  &  $-$42~~~ &   20.216\e0.019  &  $-$44~~~ &                   &       \\
Mrk335              & 108.76 & $-$41.42 & 3 &  20.48$_{-0.04}^{+0.04}$  &                & $-$100~ &  25~~~~ &  20.496\e0.016  &   $-$8~~~ &   20.555\e0.015  &  $-$10~~~ &  20.477\e0.015  &  $-$7~~~\\
Mrk478              &  59.24 &  65.03 & 2 &  19.87$_{-0.02}^{+0.02}$  &                &  $-$75~ &  50~~~~ &  19.916\e0.037  &   $-$9~~~ &   19.913\e0.040  &   $-$6~~~ &  19.927\e0.035  &  $-$9~~~\\
Mrk509              &  35.97 & $-$29.86 & 2 &  20.59$_{-0.04}^{+0.04}$  &                &  $-$60~ & 105~~~~ &  20.613\e0.013  &    7~~~ &   20.617\e0.013  &    6~~~ &  20.583\e0.012  &   8~~~\\
Mrk771              & 269.44 &  81.74 & 4 &  20.28$_{-0.03}^{+0.02}$  &                &  $-$60~ &  60~~~~ &  20.415\e0.016  &  $-$11~~~ &   20.357\e0.016  &  $-$11~~~ &  20.327\e0.016  & $-$13~~~\\
Mrk876              &  98.27 &  40.38 & 2 &  20.36$_{-0.04}^{+0.04}$  &  0@$-$139        & $-$210~ &  60~~~~ &  20.346\e0.015  &  $-$25~~~ &   20.424\e0.013  &  $-$32~~~ &  20.395\e0.015  & $-$32~~~\\
                    &   0.00 &   0.00 & 0 &  20.34$_{-0.05}^{+0.03}$  &  12@$-$139\\
                    &   0.00 &   0.00 & 0 &  20.29$_{-0.04}^{+0.05}$  &  25@$-$139\\
                    &   0.00 &   0.00 & 0 &  20.26$_{-0.04}^{+0.06}$  &  37@$-$139\\
Mrk926              &  64.09 & $-$58.76 & 1 &  20.41$_{-0.08}^{+0.15}$  &                &  $-$95~ &  60~~~~ &  20.413\e0.015  &   $-$6~~~ &   20.417\e0.015  &   $-$8~~~ &                   &       \\
Mrk1044             & 179.69 & $-$60.48 & 2 &  20.44$_{-0.05}^{+0.04}$  &                &  $-$65~ &  40~~~~ &  20.471\e0.014  &   $-$8~~~ &   20.452\e0.014  &   $-$8~~~ &  20.434\e0.014  &  $-$7~~~\\
Mrk1383             & 349.22 &  55.13 & 4 &  20.34$_{-0.04}^{+0.05}$  &                &  $-$65~ &  40~~~~ &  20.387\e0.017  &   $-$5~~~ &   20.382\e0.017  &   $-$4~~~ &  20.370\e0.017  &  $-$3~~~\\
Mrk1513             &  63.67 & $-$29.07 & 3 &  20.54$_{-0.02}^{+0.02}$  &                &  $-$75~ &  50~~~~ &  20.550\e0.013  &    2~~~ &   20.544\e0.013  &    3~~~ &  20.534\e0.013  &   3~~~\\
NGC985              & 180.84 & $-$59.49 & 3 &  20.50$_{-0.04}^{+0.04}$  &                &  $-$50~ &  50~~~~ &  20.537\e0.015  &   $-$8~~~ &   20.546\e0.014  &   $-$6~~~ &  20.534\e0.015  &  $-$7~~~\\
NGC5548             &  31.96 &  70.50 & 3 &  20.12$_{-0.02}^{+0.03}$  &                &  $-$75~ &  35~~~~ &  20.131\e0.024  &   $-$7~~~ &   20.172\e0.022  &   $-$6~~~ &  20.164\e0.021  &  $-$8~~~\\
NGC7469             &  83.10 & $-$45.47 & 3 &  20.61$_{-0.04}^{+0.06}$  &                &  $-$70~ &  50~~~~ &  20.654\e0.023  &   $-$4~~~ &   20.643\e0.021  &   $-$3~~~ &  20.637\e0.022  &  $-$3~~~\\
PG0804+761          & 138.28 &  31.03 & 3 &  20.53$_{-0.03}^{+0.03}$  &                & $-$100~ &  50~~~~ &  20.497\e0.016  &   $-$6~~~ &   20.525\e0.016  &   $-$6~~~ &  20.514\e0.017  &  $-$5~~~\\
PG0953+414          & 179.79 &  51.71 & 3 &  20.03$_{-0.03}^{+0.03}$  &                & $-$100~ &  45~~~~ &  19.997\e0.031  &  $-$14~~~ &   20.039\e0.028  &  $-$10~~~ &  19.995\e0.030  & $-$12~~~\\
PG1001+291          & 200.08 &  53.21 & 2 &  20.23$_{-0.04}^{+0.04}$  &                &  $-$90~ &  35~~~~ &  20.203\e0.020  &  $-$24~~~ &   20.210\e0.019  &  $-$22~~~ &  20.195\e0.019  & $-$22~~~\\
PG1049$-$005          & 252.28 &  49.88 & 1 &  20.68$_{-0.03}^{+0.03}$  &                &  $-$70~ &  35~~~~ &  20.584\e0.014  &  $-$13~~~ &   20.616\e0.015  &  $-$12~~~ &                   &       \\
PG1116+215          & 223.36 &  68.21 & 1 &  19.85$_{-0.05}^{+0.04}$  &  41@$-$6         &  $-$75~ & $-$25~~~~ &  19.863\e0.042  &  $-$41~~~ &   19.867\e0.041  &  $-$41~~~ &                   &       \\
                    &   0.00 &   0.00 & 0 &  19.78$_{-0.05}^{+0.05}$  &  51@$-$6\\
                    &   0.00 &   0.00 & 0 &  19.71$_{-0.07}^{+0.05}$  &  61@$-$6\\
                    &   0.00 &   0.00 & 0 &  19.75$_{-0.05}^{+0.06}$  &  54@$-$42        &  $-$25~ &  25~~~~ &  19.635\e0.068  &   $-$6~~~ &   19.679\e0.062  &   $-$6~~~ &                   &       \\
                    &   0.00 &   0.00 & 0 &  19.64$_{-0.07}^{+0.07}$  &  68@$-$42\\
                    &   0.00 &   0.00 & 0 &  19.48$_{-0.11}^{+0.09}$  &  82@$-$42\\
PG1211+143          & 267.55 &  74.31 & 3 &  20.37$_{-0.02}^{+0.03}$  &                &  $-$65~ &  35~~~~ &  20.413\e0.013  &  $-$17~~~ &   20.410\e0.015  &  $-$17~~~ &  20.413\e0.013  & $-$16~~~\\
PG1216+069          & 281.07 &  68.14 & 3 &  20.20$_{-0.04}^{+0.04}$  &                &  $-$55~ &  35~~~~ &  20.173\e0.021  &  $-$11~~~ &   20.173\e0.026  &  $-$11~~~ &  20.163\e0.021  & $-$10~~~\\
PG1259+593          & 120.56 &  58.05 & 1 &  19.75$_{-0.04}^{+0.07}$  &  (b)           &  $-$30~ &  20~~~~ &  19.437\e0.104  &   $-$6~~~ &   19.585\e0.081  &   $-$4~~~ &                   &       \\
                    &   0.00 &   0.00 & 0 &  19.65$_{-0.05}^{+0.10}$  &  (b)           &  $-$95~ & $-$30~~~~ &  19.415\e0.109  &  $-$51~~~ &   19.602\e0.078  &  $-$54~~~ &                   &       \\
                    &   0.00 &   0.00 & 0 &  19.98$_{-0.03}^{+0.04}$  &  (b)           & $-$160~ & $-$95~~~~ &  19.631\e0.070  & $-$128~~~ &   19.837\e0.047  & $-$127~~~ &                   &       \\
PG1302$-$102          & 308.59 &  52.16 & 3 &  20.43$_{-0.09}^{+0.09}$  &                &  $-$70~ &  55~~~~ &  20.497\e0.015  &   $-$4~~~ &   20.485\e0.014  &   $-$4~~~ &  20.455\e0.015  &  $-$3~~~\\
PG1341+258          &  28.71 &  78.15 & 4 &  19.88$_{-0.04}^{+0.04}$  &                &  $-$65~ &  55~~~~ &  20.002\e0.032  &   $-$9~~~ &   19.961\e0.034  &   $-$7~~~ &  19.945\e0.034  &  $-$9~~~\\
PG1351+640          & 111.89 &  52.02 & 1 &  20.46$_{-0.04}^{+0.04}$  &  0@$-$149        & $-$200~ &  50~~~~ &  20.301\e0.023  &  $-$59~~~ &   20.401\e0.021  &  $-$62~~~ &  20.419\e0.022  & $-$62~~~\\
                    &   0.00 &   0.00 & 0 &  20.39$_{-0.03}^{+0.05}$  &  33@$-$149\\
                    &   0.00 &   0.00 & 0 &  20.33$_{-0.04}^{+0.05}$  &  66@$-$149\\
                    &   0.00 &   0.00 & 0 &  20.26$_{-0.06}^{+0.04}$  &  99@$-$149\\
PG1444+407          &  69.90 &  62.72 & 2 &  20.07$_{-0.04}^{+0.04}$  &                &  $-$65~ &  35~~~~ &  20.008\e0.030  &  $-$10~~~ &   20.037\e0.028  &  $-$10~~~ &  20.022\e0.029  &  $-$9~~~\\
PHL1811             &  47.47 & $-$44.82 & 4 &  20.59$_{-0.02}^{+0.02}$  &                &  $-$45~ &  50~~~~ &  20.606\e0.018  &         &                   &         &  20.599\e0.019  &       \\
PKS0312$-$77          & 293.44 & $-$37.55 & 1 &  20.86$_{-0.06}^{+0.07}$  &                &  $-$50~ & 270~~~~ &  20.936\e0.041  &   38~~~ &                   &         &                   &       \\
PKS0405$-$12          & 204.93 & $-$41.76 & 2 &  20.55$_{-0.03}^{+0.03}$  &                &  $-$45~ &  40~~~~ &  20.508\e0.012  &    1~~~ &   20.531\e0.014  &    3~~~ &  20.509\e0.013  &   3~~~\\
PKS2005$-$489         & 350.37 & $-$32.60 & 4 &  20.52$_{-0.01}^{+0.01}$  &                &  $-$60~ &  50~~~~ &  20.575\e0.013  &    1~~~ &                   &         &                   &       \\
PKS2155$-$304         &  17.73 & $-$52.25 & 4 &  19.98$_{-0.02}^{+0.03}$  &                &  $-$55~ &  55~~~~ &  20.117\e0.024  &   $-$6~~~ &   20.092\e0.026  &   $-$4~~~ &                   &       \\
RX J1830.3+7312     & 104.04 &  27.40 & 3 &  20.70$_{-0.02}^{+0.02}$  &                & $-$120~ &  40~~~~ &  20.713\e0.011  &  $-$15~~~ &                   &         &  20.692\e0.010  & $-$16~~~\\
Ton S180            & 139.00 & $-$85.07 & 2 &  20.02$_{-0.02}^{+0.02}$  &                & $-$120~ &  30~~~~ &  20.104\e0.025  &  $-$10~~~ &   20.035\e0.055  &   $-$6~~~ &  20.031\e0.028  &  $-$6~~~\\
Ton S210            & 224.97 & $-$83.16 & 1 &  20.06$_{-0.13}^{+0.08}$  &                &  $-$50~ &  30~~~~ &  20.129\e0.023  &  $-$10~~~ &                   &         &                   &       \\
VIIZw118            & 151.36 &  25.99 & 3 &  20.51$_{-0.02}^{+0.02}$  &                &  $-$75~ &  50~~~~ &  20.543\e0.014  &   $-$3~~~ &   20.564\e0.015  &   $-$3~~~ &  20.539\e0.015  &  $-$2~~~\\
\enddata
\tablecomments{
a: Savage et al.\ (2007) report log\,$N$(\HI)=20.12$\pm$0.03 for this target.
b: For PG\,1259+593 three different components were held fixed. The table gives
the fitted result for each of the three when using the nominal values for the
other two. When varying the values for the components held fixed, the range in
the derived value for the fitted component is about $\pm$0.10 dex for the HVC,
$\pm$0.20 dex for the IVC and $\pm$0.05 dex for the LVC.
Col.\ (1): Object name. Col.\ (2,3): Longitude, latitude. Col.\ (4): Quality
factor; for complete description see Sect.~\Sresults. Col.\ (5): Logarithmic
value of $N$(\HI) derived using \Lya, with error. Col.\ (6): $N$(\HI) and
velocity of component held constant for multiple-component spectrum fittings.
See Sect.~\Sfitmethod\ for a complete description. Cols.\ (7,8): Velocity range
over which the 21-cm profile was integrated to measure $N$(\HI). Cols.\
(9)--(14): $N$(\HI) derived from 21-cm data, and brightness-temperature weighted
velocity of 21-cm cloud components from the Leiden-Argentina-Bonn, Green Bank
140-ft, and Green Bank Telescope (GBT) observations respectively. 
}
\end{deluxetable}

\subsection{Data Quality}
\par Multiple factors contribute to the reliability of the fitted \Lya\ column
densities, including the complexity of the 21-cm profile, noise in the UV
spectrum, intrinsic AGN emission, galactic and intergalactic absorption lines,
uncertainties in continuum placement, and the presence of multiple components.
Each sightline was assigned a quality flag in the range of 0 to 4. These quality
flags do {\it not} reflect the quality of the observations, but rather the
reliability with which the column densities can be measured because of structure
in the absorption and \HI\ emission profiles. The nine quality four targets have
flat continua with a high S/N ratio. The fifteen quality three targets are
generally reliable but have potential problems with continua showing mild slope
or curvature, weak emission and absorption lines, or multiple components in the
21-cm spectrum. The twelve quality two sightlines have moderately curved
continua, moderate uncertainty in continuum placement, moderately strong
emission and absorption lines, and multiple, well-separated 21-cm components of
comparable strength. The ten quality one sightlines have low S/N, possibly high
continuum curvature, strong intergalactic or intrinsic emission or absorption
lines, or multiple 21-cm components. Finally, the thirteen quality zero targets
have a significantly low S/N ratio, contain too many or too strong intrinsic
emission or absorption lines, and multiple 21-cm components. We do not use the
quality zero cases for the analysis. Generally, we do use quality one and two
results, however, since it turns out that they do not change the properties of
the distributions we calculate, but they do improve the statistics.
\par A quality factor between 0 and 4 was also assigned to each interstellar
emission component in the LAB, GB 140-ft and GBT 21-cm profiles, based on the
clarity and separation of the individual components of each profile. Many
profiles contained additional intermediate- and high-velocity (IVC and HVC)
components, which varied in their degree of separation from the low-velocity
clouds (LVC). Quality four components either are a profile with a simple
single-peak component or are components well-separated from other clouds in the
same profile. Components that merge slightly with other clouds but are mostly
separated were given quality three. Single-peak profiles with slightly extended
tails were also rated as quality three. Quality two components either are clouds
that show moderate blending with other clouds, or are components with a
single-peak profile having broad extended tails, often containing fully merged
IVC clouds. A quality factor of one was assigned for single-peak profiles that
contain multiple IVC and/or HVC clouds that mostly blend together, Quality one
was also given when the wings of clearly distinct IVC/HVC and/or LVC components
blend with the wings and peaks of other components, but the components are
clearly separate components.

In a few cases the profile was given quality zero. For instance, sometimes the
spectrum shows emission at 0 and e.g.\ $-$40\,\kms, but in one or more of the
LAB, GBT, and/or GB 140-ft spectra these blend together, so that we no longer
know how to calculate the associated column density for each. Or in one of the
spectra there appears to be a weak IVC visible, which is not seen in another
spectrum, either it is too small or because of noise. Most low-latitude
($b$$<$5\deg) sightlines were also given quality 0.

\subsection{The 21-cm to \Lya\ column density ratio}
\par In Fig.~\Fscatter\ we plot all combinations of log\,\NHI\ derived using
\Lya, Leiden-Argentina-Bonn (LAB), Green Bank 140-ft (GB) and Green Bank
Telescope (GBT) against each other. In Fig.~\Fratio\ we show the ratios of these
column densities. These figures show that although the values derived from the
21-cm and \Lya\ data are generally similar, they are not identical.
Most of the low-column density points in the scatter plots for 21-cm-only data
are values for the high-velocity clouds. Since the scatter clearly is larger for
these clouds (red and blue points) than for the low-velocity gas (black points),
they evidently have small-scale structure on scales of 9\arcmin--36\arcmin. The
low-velocity components are possibly a mixture of clouds at different distances,
each of which may have a column density similar to that of a HVC, as well as
small-scale structure. However, all of these clouds get blended together in both
the telescope beam and in the velocity-space of the sightline.

\begin{figure}\plotfiddle{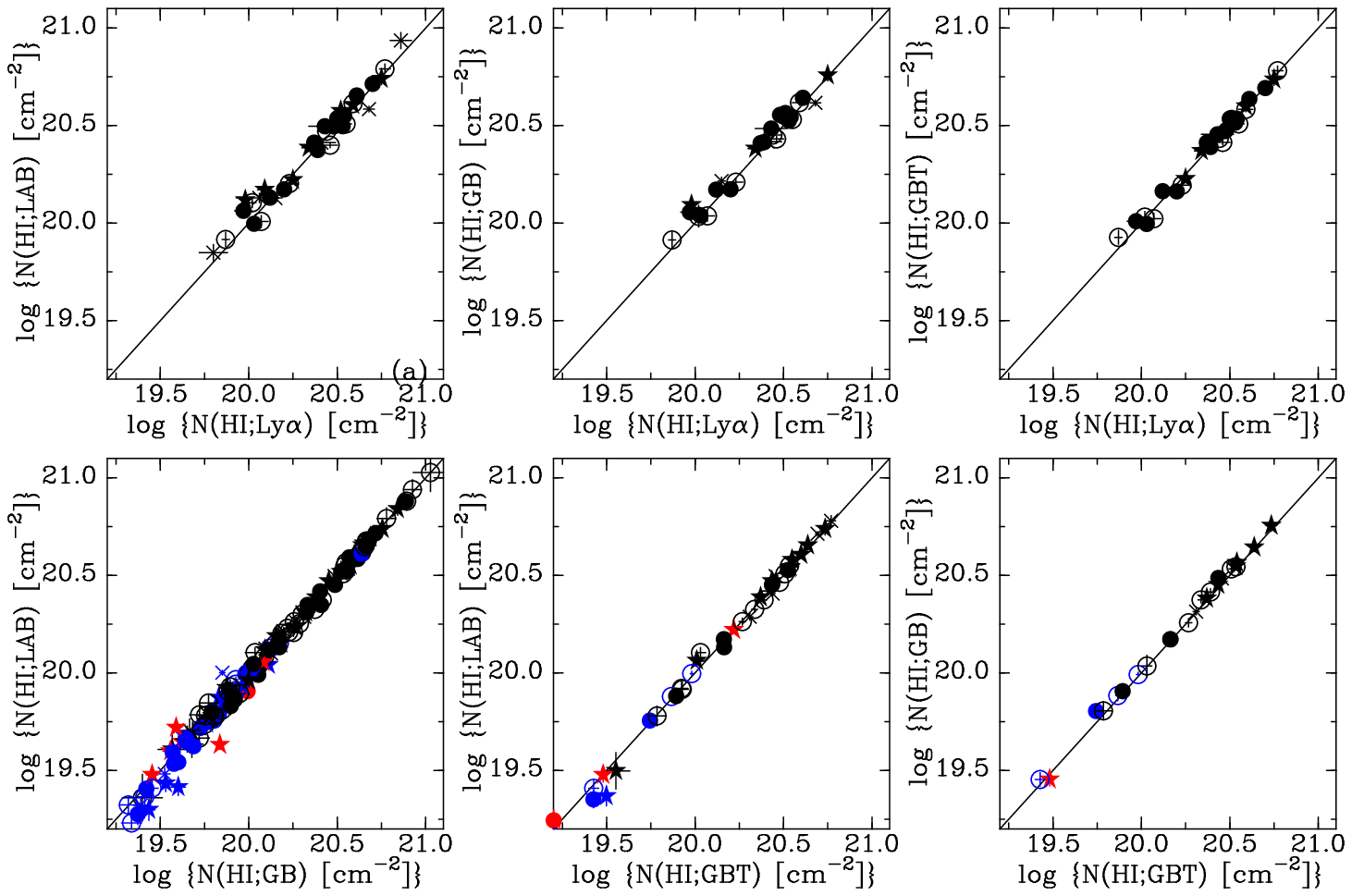}{0in}{0}{440}{300}{10}{0}\figurenum{\Fscatter}\caption{
Plot comparing column densities derived using different telescopes. All
combinations of log\,\NHI\ derived from \Lya, Leiden-Argentina-Bonn survey
(LAB), the Green Bank 140-ft (GB) and the Green Bank Telescope (GBT) are plotted
against each other. Closed stars are for measurements given quality 4, closed
circles for quality 3, open circles for quality 2 and crosses for quality 1.
Black symbols are for low-velocity gas, blue symbols for intermediate-velocity
clouds and red symbols for high-velocity clouds.
}\end{figure}

\begin{figure}\plotfiddle{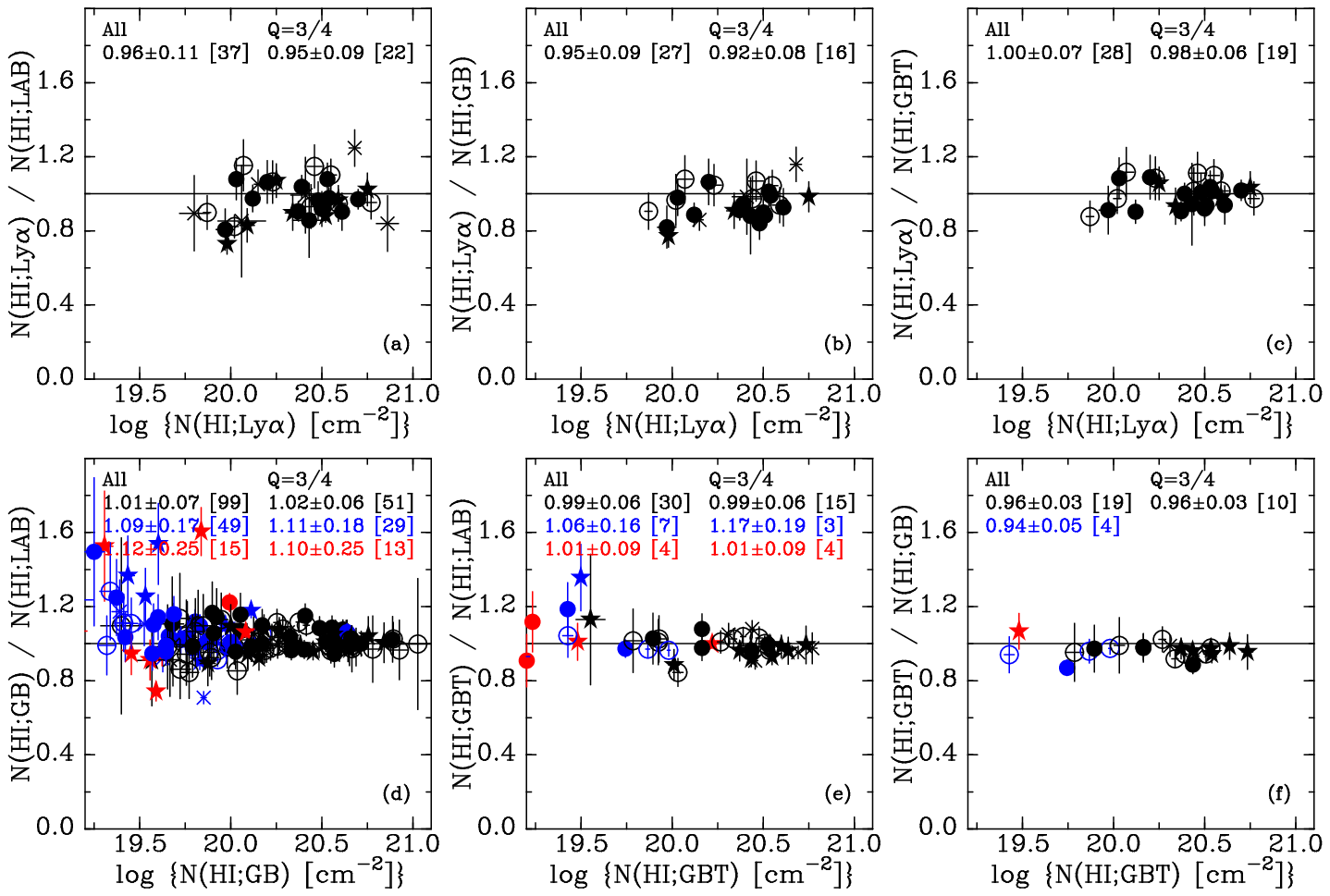}{0in}{0}{440}{300}{10}{0}\figurenum{\Fratio}\caption{
Plot showing the ratios of column densities derived using pairs of different
telescopes against the column density of one of each pair. All combinations of
log\,\NHI\ derived from \Lya, Leiden-Argentina-Bonn survey (LAB), the Green Bank
140-ft (GB) and the Green Bank Telescope (GBT) are plotted against each other.
Closed stars are for measurements given quality 4, closed circles for quality 3,
open circles for quality 2 and crosses for quality 1. Black symbols are for
low-velocity gas, blue symbols for intermediate-velocity clouds, and red symbols
for high-velocity clouds. In each of the panels the derived average ratio is
shown using either all sightlines or only the Q=3/4 sightlines, separately for
high-, intermediate- and low-velocity components, using the same color coding as
for the symbols. In addition to the average ratio, we give an estimated error in
that average (see text), as well as the number of sightlines used to derived the
average, in square brackets.
}\end{figure}

\par In the ratio panels we also give the average ratio and its rms for all
targets, and for targets with quality 3/4 only. Using LAB, GB and GBT data, on
average, the ratios for all qualities are 0.96$\pm$0.11, 0.95$\pm$0.09 and
1.00$\pm$0.07, respectively. Using only the reliable (quality 3, 4) data, the
21-cm telescopes give average ratios of 0.95$\pm$0.09, 0.92$\pm$0.08 and
0.98$\pm$0.06, respectively. See Sect.~\Sttest\ for further discussion of these
ratios. Clearly, on average the ratio is slightly less than 1 in most cases,
although not by much.

\par There are 22 sightlines for which we rated the \Lya\ measurement as high
quality (3 or 4). For these datasets the ratio \NHILya/\NHILAB\ ranges from 0.78
to 1.24, with the most extreme low ratio for toward PKS\,2155$-$304, and the
most extreme high toward PG\,1049$-$005. For 16 sightlines with Q=3 or 4 and
Green Bank 140-ft data the range is 0.77 to 1.16 (also for PKS\,2155$-$304 and
PG\,1049$-$005). Nineteen sightlines with GBT data fall in the high-quality
category, and the ratio of column densities ranges from 0.88 to 1.12 (Mrk\,478
and PG\,1444+407).

\subsection{The 21-cm to 21-cm column density ratio}
\par Figures~\Fratio d through \Fratio f show the measurements of \NHI\ when
comparing different 21-cm telescopes against each other. LVC, IVC and HVC
components are compared separately and are shown by black, blue and red symbols,
respectively. These figures indicate that the different telescopes yield similar
column densities for the low-velocity gas, which has column densities
log\,\NHI$>$19.4 to 21.0. The column densities of the IVC components span the
column density range log\,\NHI$\sim$18.6 to 20.2 (with one value near 20.6). The
HVCs span a similar range as the IVCs, but cluster toward lower values. It is
evident that the IVC and HVC column densities vary much more strongly between
different 21-cm beams than the column densities of the low-velocity gas. It is
likely that at least part of the reason for this is that the low-velocity
emission originates from multiple clouds at different distances. Consequently,
if it were possible to separate these clouds, the ratio of \NHI\ derived from
different telescopes would probably show a spread as large as that seen for the
IVCs and HVCs.

\par Comparing the LAB, GB 140-ft and GBT data to each other shows that for
high-quality low-velocity components, these three telescopes on average give the
same value for \NHI\ (average ratios of 1.01$\pm$0.07, 0.99$\pm$0.06 and
0.96$\pm$0.03, see labels in Fig.~\Fratio). The same is true for the
intermediate- and high-velocity components, although the dispersions around the
mean clearly are much larger in these cases. Also including the low-quality
components increases the dispersions, but does not noticeably change the
averages.

\subsection{Comparison with FOS results}
\par This paper was originally motivated by a desire to improve on the
comparison between 21-cm and \Lya\ \HI\ column densities using the measurements
obtained with the \FOS. In Table~\TFOS\ we show the results of Savage et al.\
(2000) for the twelve sightlines that are in both samples. In Cols.~4 and 5 we
compare the original 21-cm measurements with the ones we obtained here, showing
the decrease due to correcting for the broad underlying spurious gaussian.
Columns~2 and 3 give the \Lya\ column densities. All but one of our values are,
on average, 15\% (0.06~dex) larger than the ones obtained by Savage et al.\
(2000), as shown in Col.~6. This average excludes the factor two discrepancy for
HS\,0624+6907, whose FOS spectrum was difficult to measure, as is also shown by
the fact that for this sightline the ratio \NHILya/\NHITWcm\ was 0.46. With our
new measurement, a more reasonable ratio of 0.97 is found.
\par For this set of sightlines, the ratio \NHILya/\NHITWcm\ derived from the
FOS data was 0.81$\pm$0.09, while with the new measurements it is 1.02$\pm$0.08.

Thus, where previously we found a large discrepancy in the average, now we find
none. This is due to a combination of having much better UV data (increasing the
derived \NHILya\ by 0.06 dex, which is less than the errors in the \FOS\
measurements), and finding that the 140-ft data needed a correction (decreasing
\NHIGB\ on average by 0.05 dex). The change in the \Lya\ column densities is
probably caused by the fact that for measuring the \FOS\ spectra we needed to
model and remove strong geocoronal emission, and the number of pixels to which
the continuum could be fit was small. Nevertheless, the new values are all
within the typical quoted error (0.2--0.3~dex) of the original ones. However,
the correction in the 140-ft column densities is as important, and those
corrections are larger than the errors in each individual measurement.

\def\NH{$N$(HI)}
\begin{deluxetable}{lrrrrrrr}
\tablenum{3} \tablewidth{0pt}
\tablecolumns{7}
\tablecaption{Comparison between FOS and STIS measurements}
\rotate
\tablehead{%
\colhead{Object}&\colhead{log(\NH)$^a$} &\colhead{log(\NH)$^a$} &\colhead{log(\NH)$^b$} &\colhead{log(\NH)$^b$} &\colhead{ratio$^c$}  &\colhead{ratio$^c$}  &\colhead{ratio$^c$} \\
                &\colhead{(\Lya)}       &\colhead{(\Lya)}       &\colhead{(GB)}         &\colhead{(GB)}         &\colhead{\Lya/21-cm} &\colhead{\Lya/21-cm} &\colhead{new/old}   \\
                &\colhead{[\cmm2]}      &\colhead{[\cmm2]}      &\colhead{[\cmm2]}      &\colhead{[\cmm2]}      &                     &                     &                    \\
                &\colhead{(FOS)}        &\colhead{(STIS)}       &\colhead{(old)}        &\colhead{(new)}        &\colhead{(old)}      &\colhead{(new)}      &                    \\
\colhead{(1)}   &\colhead{(2)}          &\colhead{(3)}          &\colhead{(4)}          &\colhead{(5)}          &\colhead{(6)}        &\colhead{(7)}        &\colhead{(8)}       \\
}
\startdata
3C249.1         & 20.39$_{-0.25}^{+0.25}$ & 20.46$_{-0.04}^{+0.03}$ & 20.46 & 20.43  & 0.85 & 1.07 & 1.17\\
3C273.0         & 20.18$_{-0.05}^{+0.05}$ & 20.25$_{-0.02}^{+0.02}$ & 20.23 & 20.18  & 0.89 & 1.16 & 1.18\\
3C351.0         & 20.25$_{-0.18}^{+0.52}$ & 20.25$_{-0.03}^{+0.03}$ & 20.31 & 20.23  & 0.87 & 1.06 & 1.00\\
H1821+643       & 20.41$_{-0.17}^{+0.19}$ & 20.50$_{-0.02}^{+0.02}$ & 20.58 & 20.54  & 0.67 & 0.92 & 1.22\\
HS0624+6907     & 20.48$_{-0.66}^{+0.48}$ & 20.77$_{-0.03}^{+0.04}$ & 20.82 & 20.78  & 0.46 & 0.97 & 1.95\\
PG0953+414      & 19.92$_{-0.05}^{+0.18}$ & 20.03$_{-0.03}^{+0.03}$ & 20.05 & 20.04  & 0.74 & 0.98 & 1.29\\
PG1001+291      & 20.17$_{-0.10}^{+0.17}$ & 20.23$_{-0.04}^{+0.04}$ & 20.27 & 20.21  & 0.81 & 1.05 & 1.14\\
PG1116+215$^d$  & 19.94$_{-0.04}^{+0.04}$ & 20.05$_{-0.05}^{+0.04}$ & 20.15 & 20.09  & 0.63 & 0.92 & 1.27\\
PG1216+069      & 20.11$_{-0.07}^{+0.31}$ & 20.20$_{-0.04}^{+0.04}$ & 20.19 & 20.17  & 0.83 & 1.06 & 1.22\\
PG1259+593$^d$  & 20.13$_{-0.07}^{+0.26}$ & 20.23$_{-0.04}^{+0.07}$ & 20.19 & 20.19  & 0.88 & 1.11 & 1.26\\
PG1302-102      & 20.41$_{-0.17}^{+0.19}$ & 20.43$_{-0.09}^{+0.09}$ & 20.52 & 20.48  & 0.78 & 0.88 & 1.04\\
PKS0405-12      & 20.53$_{-0.15}^{+0.34}$ & 20.55$_{-0.03}^{+0.03}$ & 20.57 & 20.53  & 0.91 & 1.04 & 1.05\\
\enddata
\tablecomments{
a: Cols.~2 and 3: column densities measured using \Lya\ data -- values from
Savage et al.\ (2000) based on FOS in Col.~2, values measured in this paper in
Col.~3.
b: Cols.~4 and 5: column densities measured using Green Bank 140-ft data --
values from Savage et al.\ (2000) in Col.~4, values remeasured in this paper
in Col.~5.
c: Col.~6: ratio listed in Savage et al.\ (2000). Col.~7: improved ratio
measured in this paper. Col.~8 gives the ratio between the old and new
values.
d: In these two sightlines there are multiple \HI\ components with similar
strengths. In the current paper we attempted to measure the \Lya\ column
densities separately (see Sect.~\Sresults), while for the \FOS\ data this was
not done. Thus, the comparison between the \FOS\ and \STIS\ results is not as
clear-cut as for the other sightlines.}
\end{deluxetable}

\section{Discussion}

\subsection{Does \NHILya/\NHITWcm\ differ significantly from 1?}

\begin{deluxetable}{llccrrr}
\tabletypesize{\footnotesize}
\tablenum{4} \tablewidth{0pt}
\tablecolumns{6}
\tablecaption{Statistical test results}
\tablehead{%
\colhead{$N$(\HI) source} &\colhead{Comp.} &\colhead{Qual} &\colhead{\# points} &\colhead{$<$ratio$>$} &\colhead{$t$} &\colhead{$P$} \\
\colhead{(1)}             &\colhead{(2)}   &\colhead{(3)}  &\colhead{(4)}       &\colhead{(5)}         &\colhead{(6)} &\colhead{(7)} \\
}
\startdata
Lya/LAB  & TOT & All &  37 & 0.96$\pm$0.15 &  1.64 &  5.51\% \\
Lya/LAB  & TOT & 3/4 &  22 & 0.95$\pm$0.13 &  1.74 &  4.76\% \\
Lya/GB   & TOT & All &  27 & 0.95$\pm$0.13 &  2.04 &  2.56\% \\
Lya/GB   & TOT & 3/4 &  16 & 0.92$\pm$0.12 &  2.66 &  0.86\% \\
Lya/GBT  & TOT & All &  28 & 1.00$\pm$0.11 &  0.00 &   50.\% \\
Lya/GBT  & TOT & 3/4 &  19 & 0.98$\pm$0.11 &  0.81 &  21.5\% \\
\tableline
GB/LAB   & LVC & 3/4 &  51 & 1.02$\pm$0.12 &  1.22 &  11.3\% \\
GBT/LAB  & LVC & 3/4 &  15 & 0.99$\pm$0.10 &  0.39 &  35.2\% \\
GBT/GB   & LVC & 3/4 &  10 & 0.96$\pm$0.07 &  1.89 &   4.4\% \\
\enddata
\tablecomments{
Description of columns:
Col.~1: The two sources of \HI\ column densities for which the ratios toward
individual sightlines are averaged. Only sightlines with quality 3 or 4 data
have been used.
Col.~2: This column refers to the 21-cm components that are combined to derive
the column densities. ``TOT'' is used for cases with \Lya, ``LVC'' (referring to
the low-velocity gas) is used for most 21-cm telescope combinations.
For other combinations there are fewer than 5 such measurements.
Col.~3: Column showing whether only high-quality or all data were used.
Col.~4: Number of points for which a ratio could be derived.
Col.~5: The average and dispersion of the derived ratios. Note that the
dispersions are subtly different from the ones shown in Fig.~\Fratio, as they
also include the error in the ratio, rather than just giving the dispersion in
the observed ratios (see text for more explanation).
Col.~6: Student's $t$-value: $t$=(\Davg$-$$\mu(D)$)$*$$\sqrt(N)$/$\sigma(D)$
where \Davg=\Ravg$-$1 is the difference of the average ratio from 1, $\sigma(D)$
is the standard deviation of that difference, $\mu(D)$=0 corresponds to the
null-hypothesis that \Ravg=1, and $N$ is the number of points (given in
Column~3).
Col.~7: Probability $P$ that $t$ is as large as the observed value if the null
hypothesis were true, converted to a percentage. I.e., $P$ is the probability
that we find the observed value of the average ratio and its dispersion if in
reality the average ratio equals 1.
}
\end{deluxetable}

\par Although the average ratios between \Lya\ and 21-cm column densities are
close to 1, in order to formally assess whether or not they are close enough,
we used a paired $t$-test.  This is a test that compares the difference
in repeated measurements of the same sample. Using a paired $t$-test assumes
that the population is normally distributed, or at least not highly skewed, and
that the sample size is sufficiently large. For a one-tailed paired $t$-test,
the null hypothesis is that $\mu$($D$)$\geq$0, where $\mu$($D$) is the mean
difference between measurements of the population, i.e.\ it is the theoretically
expected value of the average difference. In our case, $D$ is defined as
$D$=\NHILya/\NHITWcm$-$1, and $\mu$($D$) would be 0 if in actuality the ratio
between the column densities equals 1.
\par For a paired $t$-test the $t$ value is found from:
$t$=(\Davg$-$$\mu(D)$)$*$$\sqrt(N)$/$\sigma(D)$, where \Davg\ is the average
difference calculated from the set of ratios and $\sigma(D)$ is the standard
deviation of the differences. The $t$-value is then converted to a probability,
$P$, which is the probability that, given a null hypothesis of $\mu$($D$)$\geq$0
(i.e. the expected average ratio is $\geq$1), the data randomly produces the
observed average value, \Davg. Thus, low $P$ means that the computed average is
unlikely to be observed if the null hypothesis is assumed true, which means the
null hypothesis should be rejected, i.e., the actual average ratio is not 1.
\par To test for a difference in measurements of \NHI, each sample set member is
a line of sight that has at least two different measurements of the 21-cm column
density. If there is no difference in the column densities, then one expects the
average ratio, \Ravg, to be 1. Thus, since \Davg=\Ravg$-$1, $\mu$($D$)=0. From
the data in Table~\Tres, the average ratio, \Ravg, its difference from 1, \Davg,
and the associated dispersion, $\sigma(D)$, were calculated separately for data
with quality three and four, and qualities one through four. These ratio
comparisons were done for all combinations of \Lya\ vs.\ 21-cm (LAB, GB, GBT
data) and between the different 21-cm telescopes. For the $\sigma(D)$ used to
calculate the $t$-value, we combined the dispersion in the average ratio with
the typical (i.e.\ average) error in each individual ratio. Typically, we find
that the dispersion in the ratios is comparable to, but generally larger than
the typical error in each individual ratio. Thus, if we had more accurate
measurements of all the column densities, the $\sigma(D)$ used to derive
$t$-values would not change dramatically. For instance, in the case of the
\Lya/GBT ratio, not including the errors in the ratio gives an average ratio of
1.00 with a dispersion of 0.07 (see Fig.~\Fratio c), while including the errors
in the individual ratios increases the dispersion to 0.11.

\par Table~\Tstat\ presents the probabilities for the applicable cases. This
shows that the difference between the \Lya\ and 21-cm column densities is indeed
generally not significant, with probabilities that the difference from 1 in the
average ratio is due to chance $>$3\%, except for the case of Q=3 or 4
\Lya/140-ft ratios. However, there may still be some residual systematic effects
in the 140-ft data.

\subsection{Comparing the observed distribution of column densities to models}
\par In Fig.~\Ftheor\ we compare the predictions from a number of different
models with the observed distribution of column density ratios (shown as the
black histogram). As we describe below, we used three different approaches to
make these predictions: 1) modeling the structure using a simple hierarchical
approach. 2) assuming that the distribution is log-normal, 3) using the results
of a 3-D MHD simulation.

\subsubsection{Modeling the structure using a simple hierarchical approach}
\par One way to represent the distribution of small-scale structure in column
density measurements is a hierarchical model. This means that higher column
density regions are enclosed by and cover some fraction of the area of lower
column density regions. We can construct such a model by starting with assuming
that a patch has some column density, $N$, which covers some fraction $A$ of the
area inside the 21-cm beam. Next, we assume that there are one or more other
patches with a total area that is a fraction $\beta$ times smaller, and which
have a column density that is a factor $\alpha$ larger. We then construct an
indefinite number of patches in this fashion. In the end, each patch is meant to
represent the total area covered by all cloud structures at a certain column
density within a region, regardless of where the structure is located. This
model assumes that as area decreases, column density decreases, thus $\alpha$
must be greater than 1 and $\beta$ must be less than 1. In fact, we find that
their product also needs to be $<$1.
\par We can now derive a mathematical prediction for the distribution of column
densities that can be compared to the data. We start with deriving \Nav, which
is the area-weighted column density average in a large region. Since \Nav\ is
the total column  density divided by the total area, \Nav\ for an area divided
into $k$ patches is: $$
\bar N = { A N\ +\ \beta A\, \alpha N\ +\ \beta^2 A\, \alpha^2 N\ +\ ...\ \beta^k A\, \alpha^k N \over A\ +\ \beta A\ +\ \beta^2 A\ +\ ...\ +\ \beta^k A}\ = { {1\over1-\alpha\beta} \over {1\over1-\beta} },
$$ where $A$ is the area of the first patch, having column density $N$. Thus,
$A$ can be divided out, and by extracting $N$ and inverting the formula to
derive \NoNav, and plugging in the result of the infinite-series summation, we
find $$
r_0\ =\ {N\over\bar N}\ =\ {1 - \alpha \beta \over 1 - \beta}.
$$ Here $r_0$ is the ratio of the column density in the first patch to the
average over the whole target. Each subsequent $m$-th patch will have area
$A_m=A \beta^m$, and column density ratio $r_m = r_0\ \alpha^m$.
\par To derive the distribution $f(r)$, we need to calculate the relative
fraction of the total area covered by patch $m$:$$
f(r_m)\ =\ { A_m \over \Sigma_{m=0}^{m=\infty}\ A_n}\ =\ \beta^m\ (1-\beta).$$
\par Thus, given $\alpha$\ and $\beta$, the probability of a beam hitting a
patch can be calculated, as well as the column density ratio corresponding to
that patch. The probability is the fraction of area that a single patch covers
(i.e.\ f($r_m$)). The column density ratio in that patch is given by $r_m$. This
probability can be interpreted as the number of times one would expect to
observe a beam hitting the respective patch, given a number of pencil beam
measurements taken.

\par In Fig.~\Ftheor\ we display the hierarchical model (blue curves) for
$\alpha$=1.14 and $\beta$=0.44. These values were determined such that the
modelled hierarchical distribution has similar peak and dispersion as the
observed distributions for \NHILya/\NHILAB\ and \NHILya/\NHIGBT. This
combination of parameters implies that 56\% of the area is covered by pencil
beams having a column density that is 89\% of the average, 25\% is covered by
pencil beams with column density that is 101\% of the average, while 19\% of the
beams are 116\% of the average or higher. Thus, the hierarchical model predicts
a high peak near $r$$\sim$0.90, and a small fraction of sightlines with ratio
$>$1. Observationally, for the comparison of \Lya\ to LAB column densities, we
find 49\% (18 of 37) of the beams have $r$ between 0.8 and 0.95, while 41\% (15
of 37) have $r$ between 0.95 and 1.08, and 4 (11\%) have $r$ above 1.08. Thus,
the column densities inside a 36\arcmin\ LAB beam have a distribution that is
not too dissimilar from the hierarchical model. In contrast, comparing \Lya\ to
GBT column densities yields 36\% (10 of 28) with $r$=0.8--0.95, and 43\% (12 of
28) having $r$=0.95--1.08, with 6 (22\%) at $r$$>$1.08. In this case, the
predicted peak at lower ratio is not observed. In general, the hierarchical
model underpredicts the number of high ratios.

\begin{figure}\plotfiddle{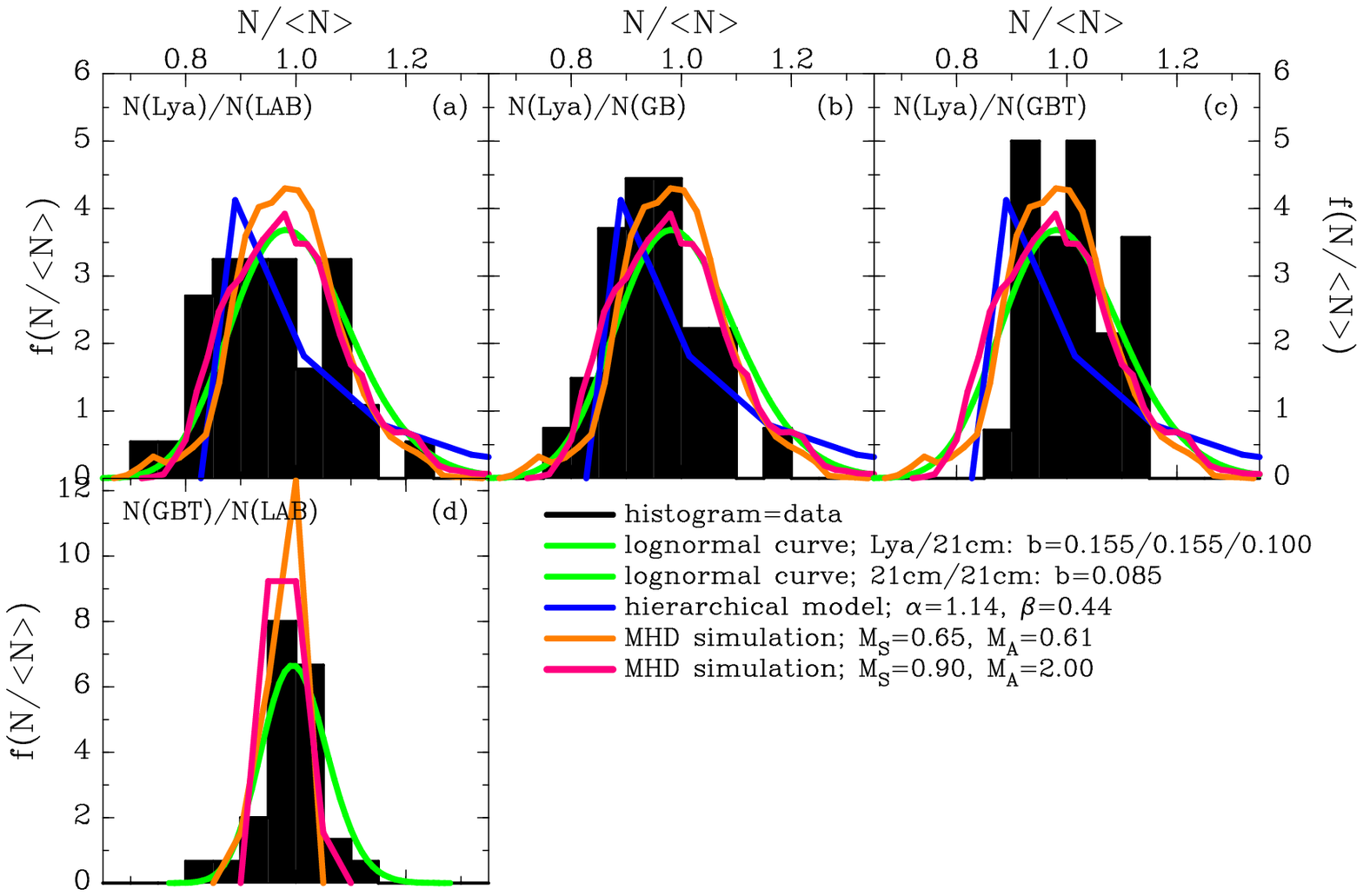}{0in}{0}{440}{300}{10}{0}\figurenum{\Ftheor}
\caption{Plot examining the distribution of \NoNav\ ratio measurements. The
horizontal axis is the ratio value and the vertical axis is the relative
frequency of each ratio, with the curves normalized so that the integral under
each curve is 1. The filled black histograms give the observed ratios for all
combinations of \NHI\ derived from \Lya, Leiden-Argentina-Bonn survey (LAB), the
Green Bank 140-ft (GB) and the Green Bank Telescope (GBT). All \Lya\ data are
used, independent of quality factor; using only Q=3/4 data does not
qualitatively change these histograms. For the 21-cm vs.\ 21-cm comparisons only
the low-velocity gas components are used. The green curve is a theoretical
log-normal distribution, with three different $b$ parameters (see text). The
blue curve is the hierarchical model with $\alpha$=1.14 and $\beta$=0.44. The
orange and red curves are the distributions obtained from taking random pencil
beam measurments in two different simulations of 3D MHD turbulence, using the
given sonic (M$_S$) and Alfvenic (M$_A$) Mach numbers.}
\end{figure}

\subsubsection{Assuming that the distribution is log-normal}
\par It is likely that the structure of Galactic \HI\ is determined by
turbulence (see e.g.\ Kowal et al.\ 2007; Burkhart et al.\ 2009). The resulting
3-D structure is then effectively determined by a multiplicative random walk.
That is, a parcel of gas will be compressed and will expand proportionally to
its current density. Intuitively, this should produce a column density
distribution that is log-normal, i.e., the log of the density has a gaussian
distribution around some mean. In most circumstances, if the 3-D turbulence
produces a log-normal distribution of volume densities, the 2-D column density
projection will also be log-normal.
\par We note that converting a log-normal distribution to a linear scale results
in the mode being at a value below the average. Taking a random sampling of
sightlines through the gas gives a distribution with the mode at some density.
Averaging this same parcel of gas by observing it with a large beam will thus
result in a value slightly larger than the mode. The offset is determined by the
width of the distribution. Using $N_m$ for the mode of the distribution, we can
write:
$$ f(N/N_m) = \exp\left( -{ \left(\ln\,N/N_m\over b\right)^2 } \right),
$$ where $f(N/N_m)$ is the log-normal distribution of the column densities and
$b$ is a parameter characterizing the width of the distribution. The column
densities are normalized to the modal value, $N_m$, The average value of $N$
observed in the area of integration, \Nav\ is then found to be: $$
{\bar N\over N_m} = { \int N/N_m\ f(N/N_m) dN \over \int f(N/N_m) dN }.
$$ To compare this to \NHILya/\NHITWcm, we have to invert this, since \NHILya\
corresponds to $N$, the column density in each pencil beam inside the area of
averaging, while \NHITWcm\ corresponds to \Nav. Then the integral works out to
be: $$
{N_m \over \bar N} = \exp{ -3 b^2\over4 }.
$$ Thus, taking for instance $b$=0.155, the ratio of mode to average is 0.98,
while for $b$=0.25 it is 0.95.

\par We have too few sightlines to determine whether the mode of the
distribution of ratios differs significantly from 1. However, we can match the
dispersion of the log-normal distribution to the observations. For the \Lya/LAB,
\Lya/140-ft and \Lya/GBT ratios this requires $b$=0.155, 0.126 and 0.101,
respectively. For GBT/LAB, 140-ft/LAB and GBT/140-ft the required $b$=0.085,
0.094 and 0.043. The green curves in Fig.~\Ftheor\ show two log-normal
distributions. For the \Lya\ to LAB and 140-ft comparisons we used $b$=0.155,
while for comparing \Lya/GBT we used $b$=0.100, the values for which the
dispersions match. In the case of comparing 21-cm to 21-cm data, we used
$b$=0.085, which matches the GBT/LAB dispersion. Clearly, a log-normal
distribution resembles the observed distributions. It is unclear whether it is
significant that the distribution of \Lya/LAB ratios is wider than that of
\Lya/GBT ratios, but it might in principle be possible to attribute this to the
fact that the GBT beam samples a smaller region, so that the relative
fluctuations are smaller.

\subsubsection{Using the results of a 3D MHD simulation}

\par Kowal et al.\ (2007) presented a set of simulations of turbulence. We used
the model datacubes that they produced to extract column densities by projecting
these cubes onto one of their faces. We added to this some cubes made with the
same software, but which were not included in their paper. These cubes can be
projected onto one axis and converted to a table of 65536 column densities. We
then plotted the distribution of these column densities, normalizing by the
average of all column densities (\Nav) in the simulation. Kowal et al.\ (2007)
used the gas and magnetic Mach numbers to parametrize their simulations. In
fact, there is a distribution of Mach numbers throughout their cubes, and the
numbers represent the average Mach numbers.
\par We extracted the predicted column density distributions for each of 16
simulations, and compared them to the data. In Fig.~\Ftheor\ the orange and red
curves give the theoretical distributions of ratios, \NoNav, for the two best
models, which are the cases with sonic and Alfvenic Mach numbers of
($M_S$,$M_A$)=(0.65,0.61) and ($M_S$,$M_A$)=(0.90,2.00), respectively. For
models with $M_S$ below 0.5 the width of the distribution is much narrower than
observed, while if $M_S$ is larger than 1.4 the predicted distribution is much
wider than observed (and the mode lies near a ratio of 0.7 when $M_S$$>$2). In
all models, the influence of the Alfvenic Mach number is modest, changing only
the details of the distribution but not the width, although the models with
higher $M_A$ fit better than those with lower $M_A$.  It is also clear that
these instances of the MHD models predict distributions that are quite similar
to a log-normal distribution with $b$$\sim$0.155.
\par The simulations were also used to compare the GBT vs.\ LAB telescope pairs.
In this cases, $N$ is the column density in an area that is
$(10/36)^2$$\sim$(1/13)th times the full size of the simulation. Since none of
the 21-cm telescopes actually make a pencil beam measurement, it is incorrect to
use each pencil beam measurement as $N$ in these instances. When comparing,
e.g., Green Bank 140-ft to LAB data, the number of 140-ft beams inside the LAB
beam is only $(36/21)^2$$\sim$3. This means that in such cases, there is no
clear difference between $N$ and \Nav, making \NoNav\ a trivial comparison
between 21-cm telescopes of similar beam size. Thus, we only compare the GBT to
the LAB data (Fig.~\Ftheor d). Using our approach the predicted width of the
distribution is clearly narrower than was the case for the comparison between
\Lya\ and 21-cm, as is predicted by the simulations.

\par Our approach of comparing column densities between observations made with
very different resolutions suggests a simple way of characterizing the
small-scale structure of the ISM. The predictions from different 3-D turbulent
MHD models are clearly sensitive to the sonic Mach number that is used. The
observed distribution of ratios fits very well with those predictions. The COS
instrument on \HST\ is expected to provide the possibility of measuring \NHILya\
toward several 100 additional sightlines, strongly improving the statistics and
thus the constraints on the model parameters. Another possibility is to use
high-resolution \HI\ data, such as those that will become available when the
GALFA (see Peek \& Heiles 2008) or GASKAP surveys are complete (GASKAP is a
survey of the Galactic Plane to be executed with the ASKAP telescope, building
of which is in progress). These surveys have 3\arcmin\ and 1\arcmin\ resolution,
and thus provide a large dynamic range in resolution.

\section{Conclusions}
\par We derive the column density of neutral \HI\ from the \Lya\ line using data
from the \STIS-spectrograph on \HST, and from the 21-cm line using the
Leiden-Argentina-Bonn (LAB) survey, Green Bank 140-ft and Green Bank Telescope
(GBT) observations. Using these column densities, we compare the ratio of
\NHILya\ to \NHITWcm\ and \NHITWcm\ to \NHITWcm\ in order to analyze the
structure of the ISM. Based on the results, we conclude the following:

\par (1) For 59 AGNs surveyed, 36 yield reliable \Lya\ column densities. There
are 163 Green Bank 140-ft and 35 GBT sightlines for which \NHITWcm\ is derived.
For each of the unique sightlines, we also measured $N$(\HI) using LAB data.

\par (2) We conclude that the published LAB data, as well as our old (from the
late 1980s) Green Bank 140-ft data, suffer from a problem that results in an
excess column density. of $\sim$1.5\tdex{19}\,\cmm2. This problem is revealed by
extracting the spectral regions outside the range where signal appears to be
present. There is no residual emission in the combined GBT spectrum, in contrast
to the LAB and 140-ft data. The residual can be fitted by a gaussian, which for
the LAB dataset has $v$=$-$22~\kms, $T$=0.048~K and FWHM=167~\kms. The
parameters of the residual are different for the 1~\kms\ and 2~\kms\ channel
spacing 140-ft spectra.

\par (3) We conclude that the \HI\ spectra from the LAB survey need to be
corrected for the presence of a broad underlying gaussian. Without such a
correction, the LAB data conflict with measurements of \NHI\ made with the GBT
and made using \Lya\ absorption, as well as with UV absorption-line studies,
and with the properties of the recently-found population of small \HI\ clouds.
With such a correction, all tension between measures of \NHI\ at different
resolutions disappears. 

\par (4) Using data from the \FOS\ on \HST, Savage et al.\ (2000) had found that
on average \NHILya/\NHITWcm\ was 0.81$\pm$0.09 for 12 sightlines. Using our new
data, the same set of sightlines yields \Lya\ column densities that are on
average $\sim$0.06~dex higher (compared to the typical error of 0.2--0.3~dex in
the \FOS\ results). We also find that a correction is needed to the 140-ft data,
which corresponds to 0.05~dex in these 12 sightlines. As a result of these
corrections, the average ratio \NHILya/\NHITWcm\ for these sightlines is now
found as 1.02$\pm$0.08.

\par (5) After applying the corrections to the LAB and 140-ft observations, we
find that the ratios between \NHILya\ and \NHITWcm\ are on average 0.96$\pm$0.11
(\Lya/LAB), 0.95$\pm$0.09 (\Lya/140-ft) and 1.00$\pm$0.07 (\Lya/GBT). A
statistical test shows that these averages do not differ from 1 in a
statistically significant way.

\par (6) A hierarchical model for the ISM matches the observed column density
distribution for \NHILya/\NHITWcm\ ratios adequately well, although it
underpredicts the number of high ratios.

\par (7) A log-normal model matches column density ratio distribution moderately
well, with a different width parameter for different cases.

\par (8) Using a 3-D MHD model from Kowal et al.\ (2007), we can match most
features of the column density distributions when choosing cases with
$M_S$=0.65--0.90. These distributions are similar to a simple log-normal
distribution, as is expected for turbulence.

\par (9) We conclude that by comparing \HI\ column densities observed at very
different resolutions it becomes possible to characterize the small-scale
structure of the ISM. Although the number of sightlines in our sample is small,
the distribution of column density ratios approximately follows a log-normal
distribution, which is also similar to the predictions of 3-D MHD modeling,
using a sonic Mach number in the range 0.65-0.90.

\bigskip
\bigskip

Acknowledgements
\par BPW acknowledges support from NSF grant AST-0607154 and NASA-ADP grant
NNX07AH42G. The Green Bank Telescope is part of the National Radio Astronomy
Observatory, a facility of the NSF operated under cooperative agreement by
Associated Universities, Inc. The \Lya\ data in this paper were obtained with
the NASA ESA {\it Hubble Space Telescope}, at the Space Telescope Science
Institute, which is operated by the Association of Universities for Research in
Astronomy, Inc.\ under NASA contract NAS5-26555. Spectra were retieved from the
Multimission Archive (MAST) at STScI. We thank UW graduate students Alex Hill
and Blakesley Burkhart for extracting the column densities from the simulation
of Kowal et al.\ (2007) and for providing additional models. JMB acknowledges
Steve Schmitt and Erin Conrad for mathematical discussions. We thank the
anonymous referee for insisting that we investigate possible errors in the 21-cm
surveys.

\bigskip
\bigskip

Appendix

\par {\it 3C\,249.1} -- The quality factor of the column density determination
is only 2, because the spectrum is relatively noisy and the 21-cm spectrum shows
the presence of several IVCs. There is an unidentified line near 1210~\AA, which
cannot be intergalactic \Lya, but which also does not fit into any known system
of absorbers toward this sightline.

\par {\it 3C\,273.0} -- Intrinsic \OVI\ emission at 1195.32~\AA\ and
1201.91~\AA\ creates a moderately broad peak in the continuum at wavelengths
shorter than Galactic \Lya. Combined with the intermediate-velocity gas at
$v$$\sim$25~\kms\, this leads us to assign quality 2 to this sightline. A third
order polynomial fits the right side of the \OVI\ wing well and creates a good
fit overall from 1197~\AA\ to 1247~\AA.

\par {\it 3C\,351.0} -- The spectrum of this AGN is flat near Galactic \Lya, and
the 21-cm spectrum is simple, resulting in quality factor 4 for this spectrum.

\par {\it H\,1821+643} -- The continuum of this target is flat and does not
contain features. However, there are weak \HI\ components at high negative
velocities ($<$$-$100~\kms), lowering the confidence in the resulting value of
\NHI\ and leading us to give quality 3 to this target.

\par {\it HE\,0226$-$4110} -- The spectrum of this AGN is flat near Galactic
\Lya, and the 21-cm spectrum is simple, resulting in quality factor 4 for this
spectrum. Where we find a value of 20.09$\pm$0.04 for the \HI\ columm density,
Savage et al.\ (2007) reported 20.12$\pm$0.03 based on an earlier analysis of
the same data.

\par {\it HE\,0340$-$2703} -- Uncertain continuum placement due to the presence
of multiple strong IGM lines results in a low quality fit for this QSO
continuum. In addition, on the short-wavelength side there appears to be an
(unidentified) intrinsic emission line, making the continuum even more
uncertain. We therefore assign Q=1 to this target.

\par {\it HE\,1029$-$1401} -- The spectrum of this AGN is flat near Galactic
\Lya, and the 21-cm spectrum is simple, resulting in quality factor 4 for this
spectrum.

\par {\it HE\,1228+0131} -- The noisy continuum of this spectrum makes it is
impossible to obtain a good fit. This is further complicated by the lack of
knowledge about the intrinsic continuum of the QSO, which is higher at
$\lambda$$<$1215~\AA\ than at $\lambda$$>$1215~\AA. Intrinsic
\SIV\ $\lambda$1062.664 (redshifted to 1186~\AA) and intrinsic
\NII\ $\lambda$1083.994 (redshifted to 1211~\AA) emission may be present, and
there is also a steep continuum drop across Galactic \Lya. These problems make
the continuum too uncertain to fit, and we do not use this sightline in our
analysis.

\par {\it HS\,0624+6907} -- The spectrum of this AGN is flat near Galactic \Lya,
but the 21-cm spectrum is not simple, resulting in quality factor 2 for this
spectrum.

\par {\it HS\,1543+5921} -- A complicated continuum combined with low S/N make a
good fit for the continuum almost impossible. In particular, \Lya\ absorption in
SBS\,1543+593 at 2800~\kms\ blends with Galactic \Lya. We do not use this
spectrum in our analysis.

\par {\it HS\,1700+6416} -- The continuum is too difficult to obtain a good fit
due to multiple IGM lines and moderate to high noise. We do not use this
spectrum in our analysis.

\par {\it MCG\,+10-16-111} -- The 21-cm spectrum shows two components of similar
strength. As explained in Sect.~\Sfitmethod, we fix each one in turn and fit the
other, in order to determine a systematic error. As can be seen in Table~\Tres,
the sum of the two stays more or less constant. As we cannot reliably compare
the 21-cm and \Lya\ column densities, we assign Q=1 to this sightline.

\par {\it MRC\,2251$-$178} -- Intrinsic \Lya\ emission peaks at 1296~\AA,
creating a moderately strong emission wing for measuring the Galactic \Lya\
absorption. A third order polynomial provides a good fit from 1195~\AA\ to
1241~\AA. Since the 21-cm spectrum is simple and the upward slope is minor, we
assign Q=3 to this sightline.

\par {\it Mrk\,110} -- Intrinsic \Lya\ emission peaks at 1259~\AA, creating an
upward slope in the continuum. A third order polynomial fits this wing
adequately and provides an adequate fit for the rest of the continuum in the
wavelength range plotted, resulting in Q=3.

\par {\it Mrk\,205} -- The 21-cm spectrum contains multiple \HI\ components
including absorption originating in NGC\,4319 at 1289~\kms\ and an HVC at
$v$=$-$204~\kms. The HVC is relatively strong, resulting in a 0.04~dex
systematic error in the value of \NHILya\ for the Galactic emission. Extra
curvature is present in the continuum and a second order polynomial was used to
handle this. On balance, however, the result is reliable and we assign Q=3.

\par {\it Mrk\,279} -- Many factors contribute to a low quality (Q=1) fit.
Multiple \HI\ components are present in the 21-cm spectrum, but only one
component at $v$=$-$40~\kms\ is used for the fitting. Strong \Lya\ emission is
also present at 1252.69~\AA\ creating a large wing in the continuum. A fourth
order polynomial is used to handle these features and fits the continuum
adequately well.

\par {\it Mrk\,335} -- Intrinsic \Lya\ emission peaks at 1247.02~\AA, creating a
large wing beginning around 1208~\AA\ and producing significant curvature in the
continuum. A fourth order polynomial fits this continuum moderately well.

\par {\it Mrk\,478} -- An order three polynomial is needed to handle the
curvature in the continuum, but the fitting lines provide a good fit over the
range of 1201~\AA\ to 1234~\AA. The curvature is caused by intrinsic \FeIII-1122
emission, centered at 1209.02~\AA.

\par {\it Mrk\,509} -- The 21-cm spectrum contains multiple components,
including an IVC at $v$=61~\kms. All components were included in the measurement
of the 21-cm and \Lya\ \HI\ column density. Strong intrinsic \Lya\ emission at
1256~\AA\ creates a large wing. A fourth order polynomial fits this wing and the
continuum well from 1180~\AA\ to 1230~\AA.

\par {\it Mrk\,771} -- The continuum is flat across the Galactic \Lya\ line.
Although the 21-cm spectrum show two strong components, their separation in
velocity is low enough that the final fit to the \Lya\ absorption results in Q=4.

\par {\it Mrk\,876} -- The 21-cm spectrum shows multiple \HI\ components,
including an HVC at $v$=$-$130~\kms, thus a systematic error in the \NHI\ value
is present due to the HVC's impact. Intrinsic \OVI\ emission at 1165~\AA\ and
1171~\AA\ also causes the continuum to slope downward at the short-wavelength
side of \Lya, but an order two polynomial still provides a good fit. These
complications lead us to assign quality factor 2 to this measurement.

\par {\it Mrk\,926} -- A third order polynomial provides a good fit to the
continuum. However, multiple factors contribute to a systematic error in the
value of \NHI. One factor is strong intrinsic \Lya\ emission at 1273~\AA\ that
creates a broad wing in the continuum. Another factor is the fact that the G140M
appears to show an extra upturn to the continuum below wavelengths of 1200~\AA\
(not easily visible in Fig.~\Fspectra). Combined with a relatively noisy
spectrum, we decided to assign Q=1 to this sightline.

\par {\it Mrk\,1044} -- Strong instrinsic \Lya\ emission at 1235.86~\AA\ adds a
large wing to the continuum. A fourth order polynomial fits this wing well and
fits the continuum adequately from 1203~\AA\ to 1223~\AA. The strong curvature
across \Lya\ results in Q=2 for this target.

\par {\it Mrk\,1383} -- An order two polynomial is needed to handle the slight
curvature in the continuum, but the fit is still good enough to derive a result
with quality factor 4.

\par {\it Mrk\,1513} -- The continuum is flat, except near 1197~\AA\ where it
shows an upturn, although there are no known intrinsic emission features near
this wavelength. The tail-end of intrinsic \Lya\ emission also causes the
continuum to rise at wavelengths above 1235~\AA. The net result of both
issues is that the continuum fit is not as reliable as it might seem,
resulting in quality 3 for this measurement.

\par {\it NGC\,985} -- Intrinsic \Lya\ emission peaks at 1268~\AA, creating a
moderate upward slope near Galactic \Lya. A third order polynomial provides a
good fit from 1195~\AA\ to 1240~\AA.

\par {\it NGC\,1705} -- The continuum is too complicated to make a good fit due
to intrinsic \NV\ emission at 1241.42~\AA\ and 1245.41~\AA\ as well as intrinsic
\Lya\ emission from the galaxy, which has a redshift of only 628~\kms. Therefore,
we do not derive \NHILya\ for this sightline.

\par {\it NGC\,3516} -- The continuum is too difficult to obtain a good fit due
to moderate to high noise and strong intrinic \Lya\ emission at 1225~\AA.
Therefore, we do not derive \NHILya\ for this sightline.

\par {\it NGC\,3783} -- The continuum is too difficult to obtain a good fit due
to the strong \Lya\ emission at 1227.50~\AA, as well as the presence of multiple
strong \HI\ components at intermediate and high velocity. Therefore, we do not
derive $N$(\HI;\Lya) for this sightline.

\par {\it NGC\,4051} -- Strong \Lya\ emission at 1218.51~\AA\ and \NV\ emission
at 1241.71~\AA\ and 1245.71~\AA\ creates a continuum too difficult to obtain a
good fit.

\par {\it NGC\,4151} -- The continuum is too difficult to make a good fit due
to strong \Lya\ emission at 1219.70~\AA\ and \NV\ emission at 1242.93~\AA\ and
1246.93~\AA.

\par {\it NGC\,5548} -- Strong \Lya\ emission at 1236.55~\AA\ adds a large wing
to the continuum. A fourth order polynomial provides an adequate fit. Since the
21-cm profile is simple, the measurement is given a final quality of 3.

\par {\it NGC\,7469} -- \Lya\ emission  at 1235.51~\AA\ creates a large rise in
the continuum, but this is fit adequately well by a fourth order polynomial over
the range of 1197~\AA\ to 1230~\AA. This results in a quality 3 measurement.

\par {\it PG\,0804+761} -- The continuum is flat from 1201~\AA\ to 1234~\AA\ but
is pushed above the fitting line at 1196~\AA\ and below at 1245~\AA. The final
fit is given quality 3.

\par {\it PG\,0953+414} -- Intrinsic \CIII\ emission is present at 1204.92~\AA\,
which adds curvature that creates a hill in the continuum over a range from
1188~\AA\ to 1236~\AA. A fourth order polynomial fits this hill well over a
range of 1205~\AA\ to 1248~\AA, but the curvature leads us to assign Q=3 to this
measurement.

\par {\it PG\,1001+291} -- The quality of this continuum is diminished by
multiple factors, including a moderate S/N ratio and the continuum resting above
the fitting line at 1183~\AA.

\par {\it PG\,1004+130} -- A high level of noise in the continuum makes it
difficult to obtain a good quality fit and a reliable value of \NHI\ for this
QSO.

\par {\it PG\,1049$-$005} -- The continuum is flat, but the high level of noise
greatly reduces the quality of the fit. As a result, we assign Q=1 to this
sightline.

\par {\it PG\,1103$-$006} -- Low S/N makes obtaining a reliable fit impossible
for this target.

\par {\it PG\,1116+215} -- The intrinsic \OVI\ emission lines are redshifted
to 1214.06~\AA\ and 1220.75~\AA, which results in a large bump in the continuum
across the Galactic \Lya\ line. This can be modeled by using a fourth order
polynomial, which fits the continuum adequately well from 1180~\AA\ to 1248~\AA.
The 21-cm spectrum shows two components of similar strength at $-$39 and
$-$5~\kms. These factors combine to yield Q=1 for the resulting \Lya\ column
density.

\par {\it PG\,1149$-$110} -- Intrinsic \Lya\ emission peaks at 1275.24~\AA\ and
a low S/N ratio make a good fit for the continuum too difficult to obtain.

\par {\it PG\,1211+143} -- Intrinsic \FeIII\ $\lambda$1122.52 emission is
redshifted to 1212.77~\AA, which pushes the continuum slightly upward on the
short-wavelength side of Galactic \Lya. This lowers the quality of the fit,
although a fourth order polynomial is used and fits the continuum adequately
well from 1180~\AA\ to 1248~\AA. The final fit quality is assigned a value
of 3.

\par {\it PG\,1259+593} -- Multiple \HI\ components are present in the continuum
including an HVC at $v$=$-$127~\kms\ and an IVC at $v$=$-$52~\kms. A two-sided
fit is used, as described in Sect.~\Sfitmethod.

\par {\it PG\,1302$-$102} -- The continuum is flat, but the low S/N visibly
diminishes the quality of the fit.

\par {\it PG\,1341+258} -- The continuum is slightly above the fitting line from
1223~\AA\ to 1230~\AA, but a linear fit still provides a high quality (Q=4) fit.

\par {\it PG\,1351+640} - The continuum contains multiple absorption features,
lowering the quality of the fit. The 21-cm spectrum shows multiple \HI\
components including an HVC at $v$=$-$156~\kms. Thus, a systematic error is
introduced into the value for \NHI\ due to the HVC's impact on the continuum.
Intrinsic \FeIII\ emission is also present at 1221.53~\AA. Although a fourth
order polynomial is used to deal with these features and provides an adequate
fit for the continuum, the uncertainty associated with the HVC's column density
is such that we assign a final quality factor of 1.

\par {\it PG\,1444+407} -- A flat continuum gives an acceptable fit. However, it
is a little above the fitting line from 1180~\AA\ to 1200~\AA\ and from
1233~\AA\ to 1237~\AA, thus reducing the quality of this fit. Combined with the
multiple 21-cm components, the final column density value is quality 2.

\par {\it PHL\,1811} -- A flat continuum is a good fit despite the presence
of intrinsic \OVI\ emission at 1230.06~\AA\ and 1236.84~\AA, which pushes the
continuum slightly above the fitting line from 1224~\AA\ to 1236~\AA. Combined
with the simplicity of the 21-cm profile, the final quality for this sightline
is 4.

\par {\it PKS\,0312$-$77} -- The 21-cm spectrum contains multiple \HI\
components, including the Magellanic Bridge at $v$=191~\kms\ and $v$=166~\kms. A
two-sided fit is used as described in Sect.~\Sfitmethod. Lehner et al.\ (2008)
gives \Lya-derived column densities of 20.78$\pm$0.06 and 20.12$\pm$0.30 for
components at 5 and 210~\kms, respectively. The combined value is 20.86. From
the LAB data, we find a column densities of 20.83 and 20.23 for these two
components. Varying the Magellanic Bridge component between 20.14 and 20.31
results in values for the low-velocity column density of 20.67$\pm$0.18 to
20.71$\pm$0.17. A combined fit yield values between 20.79$\pm$0.14 and
20.87$\pm$0.11, depending on the precise selections. Thus, we find a column
density for the Magellanic Bridge component that is 0.1 dex higher than that of
Lehner et al, but within their error. We also find that Lehner et al.\
underestimated the error for the low-velocity gas by a factor $\sim$3.

\par {\it PKS\,0405$-$12} -- The continuum is flat but a few extra features
reduce the quality of the fit. There may be \NeVIII\ emission at 1211.06~\AA\ and
intrinsic \OIV\ emission at 1238.75~\AA. \NIII\ emission at 1200.42~\AA\
pushes the continuum up slightly, and a dip from 1243~\AA\ to 1255~\AA\ pulls the
continuum down, creating a twist in the continuum that causes the fitting line to
slope upward. In the end, we only assign Q=2 to this sightline.

\par {\it PKS\,2005$-$489} -- This spectrum is flat across Galactic \Lya, and
the sightline has a simple 21-cm profile. The derived \Lya\ column density is
quality four.

\par {\it PKS\,2155$-$304} -- This spectrum is flat across Galactic \Lya, and
the sightline has a simple 21-cm profile. The derived \Lya\ column density is
quality four.

\par {\it RX\,J0100.4$-$5113} -- The continuum placement is too uncertain for
this QSO to obtain a good fit or a reliable value for \NHI. The continuum is
further complicated by the presence of an HVC at $v$=92~\kms.

\par {\it RX\,J1830.3+7312} -- The moderate curvature of the continuum reduces
the quality of the fit to Q=3, but a third order polynomial still provides a good
fit for this QSO.

\par {\it Ton\,S180} -- The fit is of lower quality due to significant curvature
in the continuum. This results from the intrinsic \Lya\ line centered at 1291~\AA,
and a rise toward the lower wavelength edge that is seen in many G140M spectra,
which is probably due to a calibration problem.

\par {\it Ton\,S210} -- The fit is of lower quality due to multiple factors. One
is the moderate S/N. Another is the uncertainty of the continuum, which rises
toward the lower wavelengths due to the wing of the intrinsic \OVI\ emission,
centered at about 1155~\AA.

\par {\it UGC\,12163} -- This spectrum is relatively noisy, and the wing of the
intrinsic \Lya\ line, centered at 1245~\AA, extends above the Galactic \Lya\
absorption, making a determination of the continuum almost impossible.

\par {\it VII\,Zw\,118} -- A flat continuum is present, but the slope combined
with a continuum dip between 1195~\AA\ to 1205~\AA\ reduces the quality of the
fit.

\end{document}